\documentclass[acmlarge,screen]{acmart}
\AtBeginDocument{%
  \providecommand\BibTeX{{%
    \normalfont B\kern-0.5em{\scshape i\kern-0.25em b}\kern-0.8em\TeX}}}

\usepackage{amsmath}

\usepackage{enumitem}
\usepackage{graphicx}
\usepackage{booktabs, makecell, multirow}
\usepackage{subcaption}
\usepackage[ruled,vlined]{algorithm2e}
\usepackage{float}
\usepackage{bbm}
\usepackage{balance}
\usepackage{soul}
\usepackage{fancyvrb}
\usepackage{xcolor,colortbl}
\usepackage{caption}
\usepackage{todonotes}
\usepackage{array}

\newcolumntype{C}[1]{>{\centering\arraybackslash}p{#1}}

\setcopyright{acmlicensed}
\acmJournal{IMWUT}
\acmYear{2022} \acmVolume{1} \acmNumber{1} \acmArticle{1} \acmMonth{1} \acmPrice{1.00}\acmDOI{11.1111/1111111}

\newcommand{\review}[1]{\textcolor{black}{#1}}
\newcommand{\new}[1]{\textcolor{black}{#1}}
\newcommand{\system}{\textit{SleepMore}\xspace}
\newcommand{\etal}{{\em et al.}}
\newcommand{\etals}{{\em et al. }}

\begin{document}
\title{\system: Inferring Sleep Duration at Scale via Multi-Device WiFi Sensing}

\author{Camellia Zakaria}
\orcid{0000-0003-4520-9783}
\affiliation{%
  \institution{University of Massachusetts Amherst}
  \country{USA}
  }
\email{nurcamellia@cs.umass.edu}

\author{Gizem Yilmaz}
\orcid{0000-0002-4387-9276}
\affiliation{%
  \institution{National University of Singapore}
  \country{Singapore}
  }
\email{gizem.yilmaz@nus.edu.sg}

\author{Priyanka Mary Mammen}
\orcid{0000-0001-7627-4479}
\affiliation{%
  \institution{University of Massachusetts Amherst}
  \country{USA}
  }
\email{pmammen@cs.umass.edu}

\author{Michael Chee}
\orcid{0000-0002-6087-0548}
\affiliation{%
  \institution{National University of Singapore}
  \country{Singapore}
  }
\email{michael.chee@nus.edu.sg}

\author{Prashant Shenoy}
\orcid{0000-0002-5435-1901}
\affiliation{%
  \institution{University of Massachusetts Amherst}
  \country{USA}
  }
\email{shenoy@cs.umass.edu}

\author{Rajesh Balan}
\orcid{0000-0001-6289-9902}
\affiliation{%
  \institution{Singapore Management University}
  \country{Singapore}
  }
\email{rajesh@smu.edu.sg}

\renewcommand{\shortauthors}{Zakaria et al.}

\begin{abstract}
The availability of commercial wearable trackers equipped with features to monitor sleep duration and quality has enabled more useful sleep health monitoring applications and analyses. However, much research has reported the challenge of long-term user retention in sleep monitoring through these modalities. Since modern Internet users own multiple mobile devices, our work explores the possibility of employing ubiquitous mobile devices and passive WiFi sensing techniques to predict sleep duration as the fundamental measure for complementing long-term sleep monitoring initiatives. In this paper, we propose \system, an accurate and easy-to-deploy sleep-tracking approach based on machine learning over the user's WiFi network activity. It first employs a semi-personalized random forest model with an infinitesimal jackknife variance estimation method to classify a user's network activity behavior into sleep and awake states per minute granularity. Through a moving average technique, the system uses these state sequences to estimate the user's nocturnal sleep period and its uncertainty rate. Uncertainty quantification enables \system to overcome the impact of noisy WiFi data that can yield large prediction errors. We validate \system using data from a month-long user study involving 46 college students and draw comparisons with the Oura Ring wearable. Beyond the college campus, we evaluate \system on non-student users of different housing profiles. Our results demonstrate that \system produces statistically indistinguishable sleep statistics from the Oura ring baseline for predictions made within a 5\% uncertainty rate. These errors range between 15-28 minutes for determining sleep time and 7-29 minutes for determining wake time, proving statistically significant improvements over prior work. Our in-depth analysis explains the sources of errors.
\end{abstract}

\setcopyright{acmlicensed}
\acmJournal{IMWUT}
\acmYear{2022} \acmVolume{6} \acmNumber{4} \acmArticle{193} \acmMonth{12} \acmPrice{15.00}\acmDOI{10.1145/3569489}



\maketitle

\section{Introduction}
\label{sec:introduction}

Sleep is essential to one's physical, emotional, and mental health. The National Sleep Foundation (NSF) recommends between 7 to 9 hours as appropriate for adults to maintain general wellness \cite{hirshkowitz2015national}. However, sleep deficiency is a common public health problem, with approximately 50 to 70 million Americans suffering from chronic sleep disorders \cite{altevogt2006sleep}. The risk factors associated with insufficient sleep are performance and cognitive deficits, and its long-term effects are correlated with severe consequences such as obesity, stress, depression, and stroke \cite{hirshkowitz2015national,altevogt2006sleep}. Thus, accurately determining sleep habits remains a topic of immense interest.

The gold standard for measuring sleep is using polysomnography (PSG), a multichannel, multimodal approach performed in controlled environments like a sleep clinic by trained technicians. These factors make PSG challenging to use for day-to-day sleep monitoring. More recently, low-cost sleep wearable devices (e.g., FitBit, Jawbone)  \cite{rosenberger201624,de2019wearable,chen2021apneadetector} have become popular for  sleep monitoring. However, wearables demand behavioral change in users putting on the device to bed, which can cause reluctance for some \cite{de2019wearable}. To overcome the challenges of continuous sleep monitoring on a longitudinal basis, researchers have explored contactless methods using radio, and radar signals \cite{rahman2015dopplesleep,hsu2017zero}. However, the instrumentation of dedicated solutions in the building infrastructure limits their support for scalability. Recently, using WiFi signals such as channel state (CSI) and backscatter information has been proposed as viable solutions for sleep tracking \cite{yu2021wifi}. This is made possible by both smartphones becoming common across most income levels \cite{pewOwnershipRace} and WiFi solutions becoming prevalent in public, institutional and residential locations \cite{pewMoreDigital,pewHomeBroadband}. These solutions detect when the user has fallen asleep by checking for vibrations and other signals. Unfortunately, they require custom hardware and are thus hard to deploy at scale, despite yielding reasonable accuracy.

Smartphones have been used for sleep detection and monitoring. Prior approaches use smartphone activity as a proxy for user activity and long periods of device inactivity to infer sleep periods. These approaches include both client-side methods, based on monitoring screen activity \cite{chen2013unobtrusive}, as well as network-side methods, based on monitoring WiFi network activity of the device \cite{mammen2021wisleep}. Since users tend to habitually use their mobile device before they sleep and upon waking up, device activity, or lack thereof, is a feasible approach for detecting sleep periods \cite{toh2019moment}. While prior work has shown the feasibility of such approaches, they are known to suffer from significant errors, often more than an hour, when determining sleep time. \new{These challenges motivate \emph{the need to develop an accurate but generally accessible sensing approach that monitor users' daily sleeping behavior without changing their routines.} Such functionality primarily contributes as a critical resource to the longitudinal requirement in almost all sleep medicine studies, facilitating complete data collection of fundamental sleep measures in real-time.}

In this paper, we present \system, a practical approach to sleep monitoring using passive observations of a user's device WiFi activity. Our approach leverages the growing number of mobile devices owned by each user in recent years (e.g., phone, tablet, laptop, e-readers) \cite{crompton2018use}. While smartphones remain the primary mobile device for most users, they may use a combination of devices over the day (e.g., use a tablet to stream content prior to bedtime and use the phone for other activities). We hypothesize that the higher errors of a single device sleep detection approach \cite{mammen2021wisleep,chen2013unobtrusive} can be minimized by observing network activity from all the user's devices to infer sleep and wake times accurately. 

\system collects network activity information of all devices directly from the WiFi access points (AP) as features. Then, it employs a two-pronged technique, running both a random forest machine learning classification and moving average estimation models to predict sleep duration. Specifically, it first classifies users in \emph{sleep} or \emph{awake} states and computes confidence intervals for these predictions using an infinitesimal jackknife variance estimation method. Outcomes with less than 95\% confidence level are noted as low confidence predictions. From applying a moving average, the most extended sequence of sleep states is observed as the user's nocturnal sleep period, with the start of the sequence denoting bedtime, $T_{sleep}$, and the end of sequence denoting wake time, $T_{wake}$. Simultaneously, the uncertainty rate of this estimation is instanced by the number of low confidence prediction states present in this sequence.

\new{Our biomedical sleep research experts} conducted an IRB-approved study over four weeks \new{of an academic semester during the COVID-19 pandemic restriction phase. Accordingly, we evaluate the performance of \system } among 46 undergraduate students who resided on campus. \new{We also conducted a small-scale user study among three non-student participants to evaluate our system's performance in home settings.} As part of the study protocols, student participants wore the \emph {Oura ring} (gen 2) \cite{oura} wearable sleep tracker for baseline. They were required to connect \new{their devices to the} campus WiFi while in their respective residences. \new{While our study specified no criteria on participants' sleep habits, all users must own multiple personal devices.} Participants were a mix of habitual and irregular sleepers. \new{They were asked to provide the MAC addresses of smartphones, laptops, and tablets} so that we could identify these devices directly from the WiFi infrastructure and extract their network event logs. By default, all WiFi traces are anonymized. \new{Our home participant chose to either provide diary logs or use their personal Fitbit. They provided their WiFi network logs directly to us.} To the best of our knowledge, this is the first work to accurately predict sleep using a scalable WiFi-based technique using inputs from multiple user-owned devices. In designing, implementing, and evaluating \system, our paper makes the following contributions:
\begin{enumerate}
    \item We present a random forest ML-based algorithm with infinitesimal jackknife variance estimation and moving average smoothing technique to predict users’ nocturnal sleep from WiFi network activity data of multiple user devices in residential spaces. Our ML classifier can predict the state of a user as sleep or awake and estimate sleep duration based on the most extended sequence of sleep states within a 24-hour period. We employ the variance estimation method to measure how confident a prediction is being made, flagging predictions beyond a 95\% confidence interval as low-confidence outcomes and calculating the uncertainty rate of sleep estimates by the number of low-confidence outcomes present in the sequence of sleep states. These techniques accurately estimate a user’s sleep duration and determine their bedtime and wake time.
    
    \begin{itemize}
        \item \system is implemented as a cloud-based web service, building a semi-personalized model that requires 40\% of users’ data for training, equivalent to 9 days of training data for 23 days of prediction.
        \item \new{Predictions made within a 5\% uncertainty rate range between 15-28 minutes of sleep error and 7-29 minutes of wake error, proving statistically significant improvements over prior work \cite{mammen2021wisleep}. Note that predictions within 5\% uncertainty rate make up 80\% of our predicted outcomes. }
        \item  \new{Our system evaluation is supplemented with results comparing \system to prior AI techniques. These comparative analyses conclude conditions under which each technique would thrive in predicting sleep using WiFi network data.}
    \end{itemize}
    
    \item We conduct an extensive experimental evaluation of \system on the student population residing on campus. Our results show that \system can accurately determine the state of users sleeping with approximately 90\% recall. These state predictions help to accurately estimate users' sleep duration, bed and wake times. 
    
    \begin{itemize}
        \item We extend our solution deployment to two residential setting. The models for home users are tested in a cross-environment, utilizing students' data as the training set. Our results demonstrate the feasibility of \system in real-world settings and with actual residents. Further, we provide insights into using smart home devices that are increasingly present in homes and utilize WiFi connections. 
        \item We characterize key factors that impact the performance of our approach, including the use of one to many devices and the lack of training data to learn night-owl sleep schedules that did not often occur among our participants.
    \end{itemize}
\end{enumerate}

\new{Conclusively, \system yields statistically indistinguishable sleep duration predictions compared to the commercialized wearable trackers. However, it is essential to emphasize that our prediction mechanism of estimating sleep duration is a partial function of the entire set of fine-grained features that Oura ring and similar wearable sleep trackers can offer. In situations where sleep studies require regular data logging over an extended period, \system is competent as a lightweight supplementary tool without requiring users to wear a sleep tracker or install a dedicated mobile app on their smartphone.}

\section{Motivation and Background}
\label{sec:backgroundMotivation}

This section provides background on sleep monitoring applications and sensing techniques, and motivates our multi-device passive-sensing approach.

\subsection{Sleep Health, a Call-to-Action}
Much work has reported sleep deprivation as a public health burden \cite{altevogt2006sleep,perry2013raising}. Researchers sought to understand the reasons and consequences of insufficient sleep in different populations by measuring standard sleep parameters such as time in bed, sleep duration, wake frequencies, and sleep latency \cite{lauderdale2006objectively,altevogt2006sleep}. These studies have noted insufficient sleep among adults as a result of lifestyles and work schedules \cite{altevogt2006sleep}, while adolescents' sleep loss is positively associated with more device use and online activities \cite{smaldone2007sleepless}. At the very least, \emph{sleep duration is documented as the most fundamental and critical predictor of different health outcomes} with longitudinal associations to weight gains \cite{magee2012longitudinal}, quality of life \cite{ohayon2013excessive}, cardiovascular illnesses \cite{cappuccio2011sleep}, cognitive impairments \cite{fernandez2020objective}, to name a few.

The paradigm shift of recognizing sleep as a critical predictor of significant health consequences \cite{hale2020sleep} has led to a fast-growing trend of consumer products offering digital sleep health options for monitoring and improving sleep. From a research perspective, it has spurred a clear call to action for clinicians to comprehensively assess sleep health among various age groups. Many of these works raised concerns over support for the basic understanding of sleep, specifically, improving the effectiveness of sleep screening \cite{perry2013raising, altevogt2006sleep} over longitudinal periods and developing new technologies to accomplish this task \cite{mitler2000sleep}.

\subsection{Sleep Sensing Technologies}
Sleep studies in clinical practice generally utilize retrospective scales, daily sleep logs, and/or polysomnography. However, it is challenging for everyday consumers to personally monitor their sleep behavior with these methods. Further, it remains highly burdensome for sleep studies to conduct long-term evaluation. Hence, sleep technologies offer a low-burden approach to automatically detect a user's sleep in the comfort of their own homes. \new{In a comprehensive study over two months, Massar \etals \cite{massar2021trait} compared and contrasted sleep measurements from a consumer sleep-tracker, smartphone-based ecological momentary assessment, and user-phone interactions of 198 users. Their investigation identified stable interindividual differences in sleep behavior, underscoring the utility of these modalities in characterizing population sleep and peri-sleep behavior. We discuss the pros and cons of these options as follows:}

\begin{enumerate}
    \item \textbf{Contact-based wearables:} Commercial alternatives have shown feasibility in monitoring sleep. Zambotti's defined \emph{sleep wearable trackers} as ``over-the-counter, relatively low-cost devices available without prescription or clinical recommendations,'' \cite{de2019wearable} varying from wristbands to smartwatches, earbuds to rings. The study protocol employing sleep wearables typically involves loaning these devices to enrolled participants over a fixed duration for reusability in future studies. The implications of such practice could result in behavior modification during the study. Specifically, users not used to wearing a device to bed and raising the challenge of a high user attrition rate \cite{massar2021trait,zhai2020making,de2019wearable}.
    
    \item \textbf{Contactless sensing:} Low compliance with wearables has motivated the design of contactless sensing approaches. Examples include the use of cameras~\cite{huang2021nose}, RF or radar sensing to characterize sleep stages~\cite{rahman2015dopplesleep} and posture~\cite{yue2020bodycompass}. Despite these advances, RF sensing systems have yet to accelerate commercial adoption, while camera-based solutions are more likely to intrude on privacy interests.
    
    \item \textbf{Smartphone-based sensing:} By contrast, smartphone-based sensing arose as an option from the developing smartphone dependency \cite{pewDeviceOwnership} among everyday users, leading to researchers using the ``phone-as-a-sensor.'' Efforts specific to sleep monitoring include distinguishing respiratory patterns through a microphone~\cite{tiron2020screening} or inferring user sleep behavior from monitoring screen interactions and application usage \cite{abdullah2014towards,cuttone2017sensiblesleep,min2014toss,hao2013isleep,gu2015sleep}. In all these cases, the solution presents a dedicated mobile application that must be installed in users' smartphones and its data acquisition running in the device's background.
\end{enumerate}

\new{These works collectively underscore the advancement in sleep monitoring technologies. While promising, much tension in utilizing the modalities above in sleep studies arises from the challenge of user retention. Massar \etals describe designing an incentive structure to continuously encourage regular data logging across such modalities over the study period. With many people operating e-devices for minutes to hours before actually intending to sleep, we seek to establish an accurate and generally accessible sensing approach that monitors users’ daily sleeping behavior without changing their routines. It is important to emphasize that our work aims not to replace the utilization of existing modalities that can obtain key physiological measures. However, it is positioned as a complementary mechanism to support sleep monitoring scenarios over longitudinal periods. }

\subsection{Passive WiFi-based Sensing} Attempts to bypass the high attrition rate in smartphone-based sensing led to the utilization of WiFi-based passive sensing techniques. Prior work by Mammen \textit{et al.} has shown that it is possible to use WiFi connection logs through users' single smartphone collected from the WiFi infrastructure to predict sleep \cite{mammen2021wisleep}, however yielding only 88.50\% accuracy. Separately, broader surveys on behavioral monitoring via smart devices have argued that utilizing single-device for monitoring is not fully comprehensive, in part because of inadequate device coverage from users owning multiple devices \cite{trivedi2020empirical,pewPromisePitfall,pewDeviceOwnership}.

\begin{enumerate}
    \item Prior work only uses data collected from a single smartphone based on the assumption that users spend most of their time online and on their smartphones \cite{van2015modeling}. We hypothesize that \textbf{including all devices owned by the user will significantly improve the accuracy of sleep detection}. 
    
    \item \new{There is a significant gap between the accuracy of smartphone-based methods and those achieved by sleep wearable trackers, which} can achieve 96\% recall at detecting sleep \cite{altini2021promise}. With this result in mind, \new{it is essential that our proposed solution, while using coarse-grained data source, achieves \textbf{no statistically significant difference from a wearable sleep tracker} in order to serve its purpose.} In this study, we use the Oura Ring (gen 2) as a representative wearable device for sleep monitoring.
\end{enumerate} 

\subsection{Design Rationale}
\label{sec:motive}

\begin{figure}[h!]\centering
  \includegraphics[width=.9\linewidth]{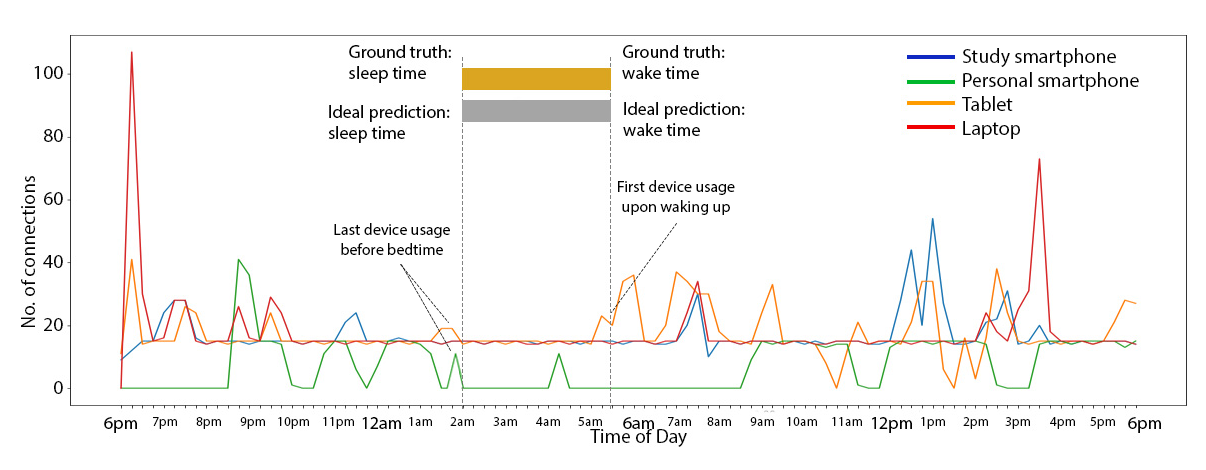}
  \vspace{-0.1in}
  \caption{Utility of smartphone and other devices before sleep and upon waking up. Low WiFi device activity from multiple devices corresponds to a sleep behavior.}
  \label{fig:rationale}
\end{figure}

Figure \ref{fig:rationale} shows an example of the network connection frequency for a typical user with multiple devices every 15-minutes through 24 hours between Day$_{\text{1}}$, 6 pm to Day$_{\text{2}}$, 6 pm. The observation that the personal smartphone is the last device used before bedtime \cite{toh2019moment} makes it viable as a sleep monitoring sensor. However, users tend to own multiple devices, and in this case, a secondary tablet device denotes the first active usage upon the user waking up. This behavior presents practical reasoning to infer users' sleep behavior more accurately through multiple device usage. Hence, this work utilizes the WiFi network device activity collected from multiple devices to estimate an individual's nocturnal sleep duration over a longitudinal basis as the most fundamental feature supporting sleep studies.

Specifically, it monitors the devices that are associated with the network to infer the user's bedtime and wake-up time (referred to as $T_{sleep}$ and  $T_{wake}$). Adding new user-owned devices is easy as they would only need to be connected to the same WiFi network. In addition, our solution significantly preserves privacy as it does not require decoding the actual content of any packets sent by a device. For each user-owned device (denoted by their MAC address), we capture the WiFi connection events that these devices make to the network and compute the frequency at which each device establishes network connections. 

The above example yields several critical insights: (1) The connection frequency increases with active online device utilization and decreases with less utility. For example, we can observe the user displaying the highest device use between 6-7 pm using a tablet. (2) A user also switches between four devices throughout the 24 hours, highlighting the potential for a more comprehensive behavioral monitoring by expanding device selection. (3) The user's smartphone is the last device used before bedtime, but it is not always the first device used upon waking up. (4) Further, network activity for other devices such as the laptop is observed to pick up soon after. (5) While the frequency of connection increases with more device use, it is important to note that increased device activity does not always imply a user being active in true nature. For example, it can occur from an application on the device running automatic software updates or accepting notifications (e.g., incoming emails, received online messages), as demonstrated by the user's smartphone network connection log, which peaked between 4 to 5 am. These observations suggest that multi-device monitoring will lead to significantly better sleep prediction than using a single device. Indeed, our results in Section \ref{sec:evaluation} confirm our hypothesis. 

It is important to emphasize that our approach aims to infer users' sleep duration by estimating their sleep and wake times, achieving robustness across users. The properties of utilizing a coarse-grained data source will innately restrict our technique in predicting more nuanced sleep characteristics such as the REM stages of sleep. Hence, our technique aims not to replace existing sleep-sensing modalities that offer fine-grained information but instead complement the use of such modalities in supporting sleep monitoring, especially over long periods.

\subsection{Key Systems Challenges} 
\label{sec:challenges}
\begin{figure}[h!]\centering
  \includegraphics[width=.9\linewidth]{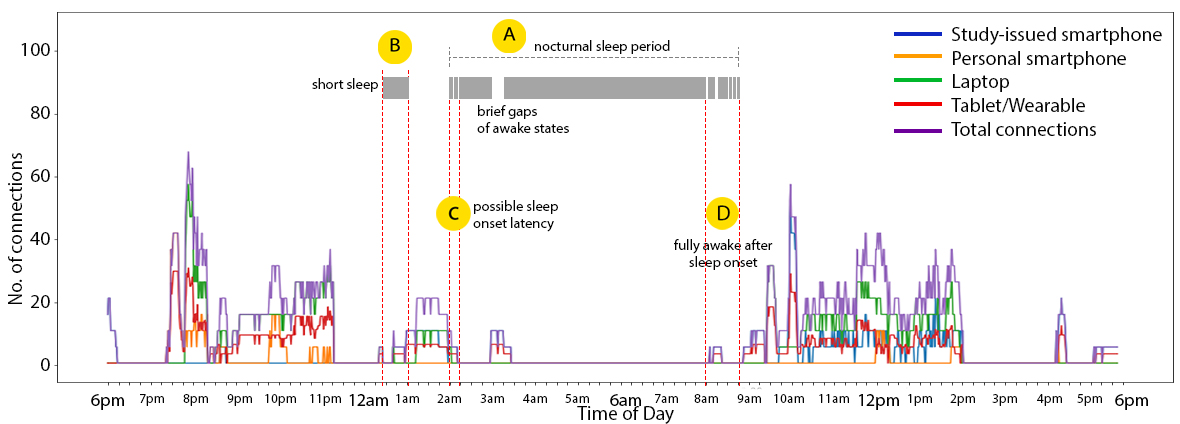}
  \vspace{-0.1in}
  \caption{Challenges predicting sleep with WiFi data.}
  \label{fig:rationale2}
\end{figure}

Our proposed approach demonstrates non-trivial challenges in using device network events as a sleep predictor, as shown in Figure \ref{fig:rationale2}. 

\subsubsection{Noisy Data}
Prior work has discussed how noise in WiFi-sensing, such as ping-ponging of AP connections, can negatively impact prediction performance \cite{mammen2021wisleep}. In our case, the AP ping-pong effect is largely minimized by limiting WiFi network data collection to within users' residential premises. However, our system will still be susceptible to other noise-affected errors. Specifically, our technique assumes that increased WiFi network activity on a user's device indicates the user is awake. It is also reasonable to assume a user interacting with different devices before bedtime and upon waking up \cite{exelmans2016bedtime,kubiszewski2014association}. However, this increased activity can be caused by device updates and app notifications that do not reflect the user's actual physical state. For example, the user is asleep at 3:00 am while the device updates or apps receive notifications (A). The variability of WiFi network activities resulting from both user and device actions will, to a large extent, affect our technique's ability to accurately predict a user's sleep duration. 

Our system is designed to overcome these challenges by estimating the confidence intervals for predicting sleep states throughout a 24-hour period. While sleep states with less than 95\% confidence level are noted as low confidence predictions, it continues to estimate the user's nocturnal sleep duration based on the longest sleep state sequence. However, it calculates the uncertainty of this estimation based on the number of low confidence predicted states present in this sequence. We provide the details to our technique in Section \ref{sec:mlmodel}.

\subsubsection{Accurate Estimate of Bed and Wake Times}
While it is not unusual for a person to use their device before bedtime, it is also not unusual for a person to take some time to fall asleep after putting their device away \cite{exelmans2016bedtime,kubiszewski2014association}. This duration is known as {\em sleep onset latency} and is estimated to be within 20 minutes. However, it also increases progressively with age and electronic device-use \cite{bartel2019altering,he2020effect,kushida2012encyclopedia}. In Figure \ref{fig:rationale2}, the user is predicted to sleep briefly between midnight and 1:00 am before going back to bed at 2:00 am (A and B). Thus, the challenge is determining the true start of a user's bedtime. Further, the first predicted sleep state may not reflect that the user is falling asleep. Conversely, the last predicted sleep state may not imply a user waking up (C and D). 

This challenge informs our decision to estimate the user's nocturnal sleep duration based on the most extended sequence of sleep states, as mentioned above. Detailed in Section \ref{sec:estmodel}, we employ moving average to smooth out short-term occurrences in wake states and highlight longer-term sleep trends over the 24-hour period.

\section{\system Design}
\label{sec:system}
\begin{figure}[h!]\centering
  \includegraphics[width=\linewidth]{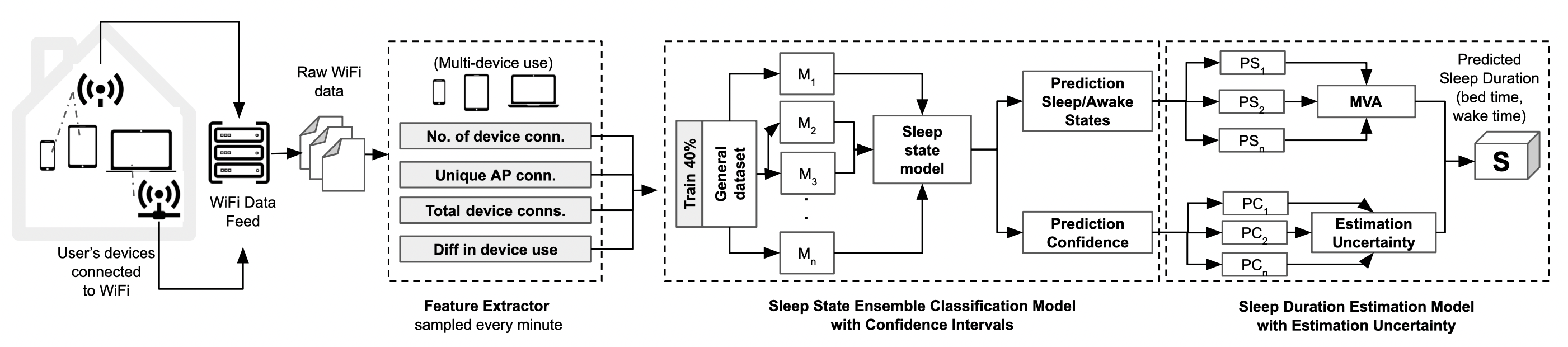}
  \caption{\system, a two-step process in predicting sleep states with confidence interval, and estimating bed and wake times.}
  \label{fig:sysOverview}
  \vspace{-0.1in}
\end{figure}

This section presents our design of \system as a multi-device WiFi sensing approach for sleep monitoring. Figure \ref{fig:sysOverview} presents the system overview. We describe the implementation details as follows.

\subsection{Collecting WiFi Data Using RTLS}
\label{sec:rtls}
\system requires WiFi data indicating when a user's device(s) communicate on the network. We leverage the real-time location system (RTLS) data feed available on commercial access points. For this paper, we collected all our data from a university dormitory using Aruba WiFi APs, which support RTLS \cite{RTLS}. The RTLS feed provides reports from every AP that sees a device communicating on the network. Each RTLS message contains the following information:

\begin{center}
    \texttt{Timestamp, Packet Age, Data Rate, DeviceType, Channel, DeviceMACAddress, \\
    AssociationStatus, RSSI, Noise Floor, BSSID, MON BSSID}
\end{center}

\system only uses data from associated client devices as those clients have legitimate access to the network. In particular, we discard all data from unassociated devices as they are primarily WiFi probe requests from clients roaming for a usable network. Note: while RTLS feeds can be used to track the location of WiFi devices, \system is only using this data as a proxy for device activity.

The RTLS reports are generated by all APs in our test environment every five seconds. However, these reports will only list the WiFi devices that were active and seen by that AP (i.e., they were using the same WiFi frequency as the AP and close enough to that AP) in that five-second interval. In particular, if a device has gone to sleep (e.g., because the user is sleeping and the device is charging), the device will not be emitting any network packets. It will not be seen and reported by any RTLS report generated by any AP.

\subsection{Pre-processing Module}
\label{sec:features}

The first step in \system's pre-processing pipeline is to clean the noisy WiFi RTLS data. In particular, we remove data with invalid timestamps or weak RSSI values indicating spurious transmission. We only retain records of multiple device WiFi activities on days that users provide their Oura data (see Table \ref{tab:datasumm} for data summary). 

\system's features are generated from the RTLS \emph{Timestamp} (records time at which a data message is received), \emph{Device Type} (client or AP station), \emph{Device MAC Address} (client or AP station), and \emph{Association status} (associated or unassociated device) fields. Because a user could own multiple devices, each with their own MAC address, \system generates device features primarily based on the number of associated connection events per device (i.e., \emph{networkEvents}) and the unique AP (i.e., \emph{uniqueAPs}) to which these devices were connected. We use the AP to which the user's device is connected as a feature for avoiding cases where a user is moving with a phone in power-saving mode -- in this case, the phone will not generate many network events, but the change of APs show that the user is not sleeping. We assume that a device connects to just a single unique AP when a user is sleeping -- this assumption can be relaxed easily (by adding a set of APs into the feature set) in the rare cases where this assumption is not true. In Section \ref{sec:frequencysensitivity},  we show that generating these features every one minute struck a good balance between accuracy, computational overhead, and resilience to noise. 

\begin{figure}[H]
\includegraphics[width=.55\columnwidth]{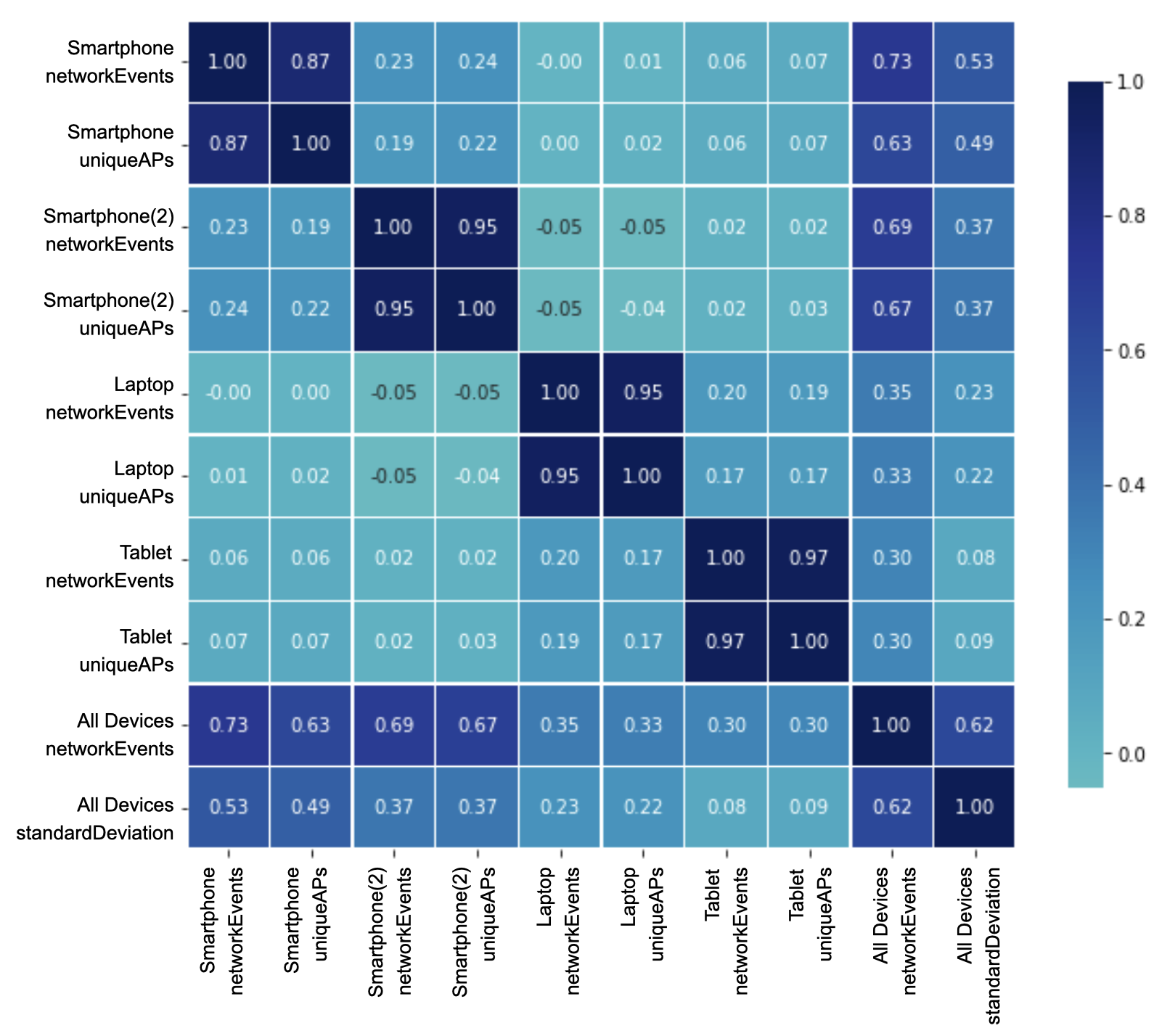}
\caption{Correlation heatmap of features extracted from multiple device WiFi network activities.}
\label{fig:featureheatmap}
\end{figure}

Figure \ref{fig:featureheatmap} shows the correlation values between sets of features (highly correlated features are darker in color and have scores close to or exactly 1.0). We observe that for all device types (smartphone, laptop, tablet), the number of unique AP associations (uniqueAPs) is highly correlated with the number of WiFi connection events (networkEvents). In Section \ref{sec:featureanalysis}, we use these results to remove all highly-correlated features (i.e., unique APs for all devices) and adopt a standard machine learning pipeline for feature selection. \system's final feature selection uses importance weights, which is the number of times a feature is used in the fitted trees inside the Random Forest classifier. As a last step in the pre-processing pipeline, it assigns sleep labels for our algorithm. We binarize the users' reported nocturnal sleep duration (i.e., Oura ring baseline) into \emph{sleep (1)} or \emph{awake (0)} states at every interval.

\subsection{Sleep Prediction Module}
\label{sec:predModule}

\begin{figure}[h!]
\includegraphics[width=\columnwidth]{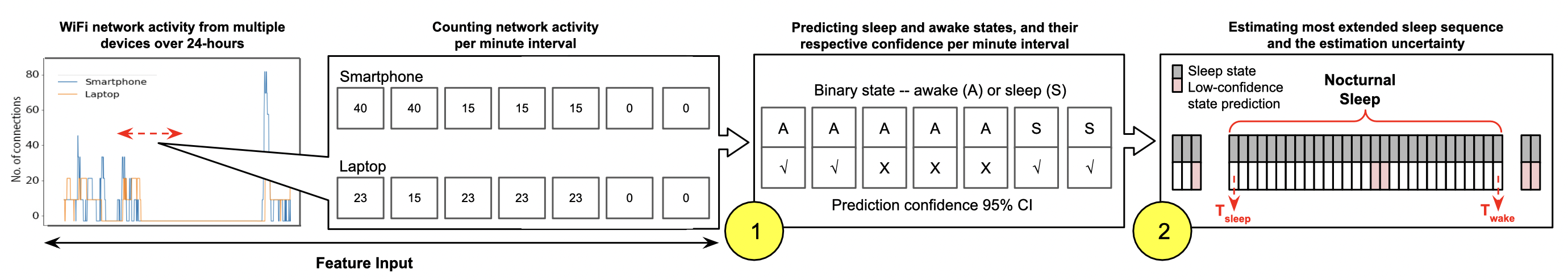}
\caption{High-level representation of our technique, receiving WiFi-generated device features as input to an ML classification and estimate sleep through sequence of sleep states.}
\label{fig:technique}
\end{figure}

\review{Figure \ref{fig:technique} provides a high-level visual representation of the inner workings of our technique. It generates a set of attributes from the user's multi-device WiFi device activity logs collected every one minute between $Day_{1}$, 6 pm and $Day_{2}$, 6 pm. Using the features as input, the system first runs a machine learning algorithm to classify a user's state as \emph{sleep} or \emph{awake}, and calculates the respective prediction confidence. Then, it takes the longest sequences of sleep states to define when the user is sleeping. In doing so, it applies a smoothing function to the sequence of events and calculates the estimation uncertainty based on the number of low-confidence states present in the sequence.}

\subsubsection{Machine Learning Classification with Confidence Interval}
\label{sec:mlmodel}

\system produces a binary predictions of \\whether the user is currently in a \emph{sleep (1)} or \emph{awake (0)} state. Section \ref{sec:resultML} compares several machine learning techniques, including Naive Bayes (NB), Random Forest (RF), Extreme Gradient Boosting (XGB), \new{and Long Short Term Memory (LSTM)} algorithms for this problem. We found that \new{the RF and LSTM yield best results.}

In RF, for \emph{b=1} to \emph{B=200} total trees, it draws a bootstrap sample from the training data. Then, it repetitively grows a tree, $T_{b}$, from the bootstrap sample by randomly selecting \emph{d=5} features without replacement and splitting the node using the feature that provides the best split until the desired node size is reached. As the algorithm outputs the total ensemble of all trees to predict a state, $\hat{f}(x)$, it aggregates the prediction by each tree to assign the class label,  $\hat{f}(x) = \frac{1}{B}\sum\limits^{B}_{b=1}T_{b}(x)$, at every one minute interval. A great advantage to using random forest is that the bagging technique helps reduce the variance of unbiased estimated prediction functions. However, our model must grow deep trees to maintain a low bias and decrease prediction variance. This complexity may lead to less interpretable results, especially in characterizing the statistical distribution of our predictions. 

\new{On the other hand, a 2-layer LSTM demonstrated comparable results to RF. The LSTM incorporates three different matrices inside the recurrence, acting as a gating mechanism that allows for information flow from the previous timestep as a function of the current one; an input gate ({$i_t$}), a forget gate ({$f_t$}) and an output gate ({$o_t$}). Here, $w$ and $b$ represent the weight and biases for gate, $x$, neurons. $h_{t-1}$ represents the output of the previous LSTM block.}

\begin{center}
$i_t = \sigma(w_i[h_{t-1},x_t] + b_i)$\\
$f_t = \sigma(w_f[h_{t-1},x_t] + b_f)$\\
$o_t = \sigma(w_o[h_{t-1},x_t] + b_o)$
\end{center}

\new{These representations in a cell state, {$c_t$} are combined with the memory vector for the current timestamp, $\tilde{c_t}$, to produce the hidden representation, {$h_t$}. Our final layer is a fully connected layer with a \emph{sigmoid} activation and one neuron. Where $W$ is the weight connection matrix of the layer, the dense layer outputs, $y_t$, the probability of the event to be of state \emph{sleep} or \emph{awake}. }

\begin{center}
$\tilde{c_t} = tanh(w_c[h_{t-1},x_t] + b_c)$\\
$c_t = f_t * c_{t-1} + i_t * \tilde{c_t}$\\
$h_t = o_t * tanh(c^t)$\\
$y_t = (h_t * W) + b$
\end{center}

\subsubsection*{Infinitesimal Jackknife Variance Estimation}
To better determine the predictions our model is more confident about, we use the infinitesimal jackknife variance estimation method proposed by Wager \etals \cite{wager2014confidence}. The method takes the covariance with respect to the resampling distribution to generate variance estimates, $\hat{V}_{IJ}$, for all predictions. That is,

\begin{center}
$\hat{V}_{IJ} = \sum\limits^{n}_{i=1} Cov_{*} [T^{*}(x), N^{*}_{i}]$
\end{center}

where $T^{*}(x)$ is the prediction, $x$, at tree, $T$, based on the subsample $M^{*}_{1}, ..., M^{*}_{s}$ of size, $s$, and $N^{*}_{i}$ is the number of times the original sample appears in the resample. This technique would down-weigh each observation by an infinitesimal amount to estimate the variance of a statistic. However, since the variance estimate is calculated with a finite number of trees in practice, it is inherently associated with Monte Carlo error. Our implementation decreases this error by using a large number of trees and subtracting off the Monte Carlo estimate of variance with bias correction. As suggested by Wager \etals \cite{wager2014confidence}, the unbiased variance estimate, $V^{B}_{IJ}$, is defined by: 

\begin{center}
$V^{B}_{IJ} = \sum\limits^{n}_{i=1}C^{2}_{i} - \frac{s(n-s)}{n} \frac{\hat{v}}{B}$
\end{center}

where \new{$C_{i} = \frac{1}{B}\sum\limits^{B}_{b=1}(N^{*}_{bi} - s/n)(T^{*}_{b} - \bar{T}^{*}) $}, given $n$ is number of training examples, and $\hat{v} =\frac{1}{B}\sum\limits^{B}_{b=1}(T^{*}_{b} - \bar{T}^{*})^{2}$. \\Finally, given the variance estimates, we find the sample mean prediction probability with a 95\% confidence interval by randomly sampling 1000 predictions per user. We labeled predictions with low confidence for those outside the upper and lower confidence interval.

\subsubsection{Sleep Duration Estimation Model}
\label{sec:estmodel}
The outcomes from our ML model are a sequence of predicted states, $S = \{s_{1}, ..., s_{m}\} $ and their respective confidence, $C = \{c_{1}, ..., c_{m}\}$, every 24 hours (\emph{m} = 1440 minutes).
By putting all binary state predictions together, we can conceivably determine a user's nocturnal sleep period for that 24-hour cycle by choosing the most extended occurrence of sleep states (Figure \ref{fig:rationale2}A). This nocturnal sleep period could include brief awake states as it is not unusual for a user to wake up once or twice during the night. However, our consideration toward the most extended sleep period, specifically during the night, may perversely overlook short sleep states that can occur between long awake states as in  Figure \ref{fig:rationale2}B.

Thus, the system component requires solving an estimation problem to accurately identify both a user's bedtime ($T_{sleep}$) and wake time ($T_{wake}$) using the series of predicted binary states. We compare two different estimation techniques for this problem: predicting with moving averages (i.e., MVA) and smoothed aggregation (i.e., AGG). Both prediction methods build on the hypothesis that a user is more likely to be predicted in a sleep state when their past events assert that they were sleeping and vice versa. 

\begin{enumerate}
    \item \textit{Predicting with Moving Averages (MVA): } When using MVA, \system will provide sufficient weight to past states in the predictor to reduce the probability of predicting an awake state when a sudden burst of WiFi activity occurs amid a nocturnal sleep period (e.g., a background app notification). Instead, it will require a series of steadily rising WiFi network device activities to change the prediction from asleep to awake. Similarly, valuing the past awake states as an additional predictor would reduce the model predicting a sleep state when WiFi device activity briefly dips (e.g., the user leaves home for a short time). 
    \item \textit{Smoothed Aggregation (AGG): } Another approach is by applying AGG whereby we total the sum of predicted sleep states over a larger observation window. For example, by considering predicted sleep states over a 30 minutes window, we produce a range of prediction states that will likely contain a 95\% confidence level of the user's state we are interested in. We smooth out the representation by applying a Savitzky-Golay (SG) filter \cite{press1990savitzky} to determine the most extended sleep period. With this smoothing, even brief sleep episodes, as shown in Figure \ref{fig:rationale2}B, will be considered as part of the user's sleep period. Then, we can determine the start of the most extended sleep period as $T_{sleep}$ and the end of the period as $T_{wake}$. 
\end{enumerate}

Our comparison on the performance of these estimation methods found MVA worked best in Section \ref{sec:estimatingSleep}. We empirically determined the sliding window, $W = 5$, based on the achieving optimal performance. For each sequence of states,  we calculate the moving average at time period, $t$, as:

\begin{center}
	$ MVA_{t} = \frac{s_{t} + s_{t-1} + s_{t-2} + ... + s_{W-(t-1)}}{W}$
\end{center}

If $s_{t}, s_{t+1}, s_{t+2}, ..., s_{T}$ is a sequence of sleep states, our model identifies the longest continuous sequence as the nocturnal sleep period for that 24-hour period. The start of a sleep period corresponds to the first occurrence of sleep states, $T_{sleep} = s_{t}$. The end of a sleep period corresponds to the last occurrence of a sleep state within the longest sequence $T_{wake} = s_{T}$. 

\subsubsection{Use of Uncertainty Quantification to Address Noisy Data}
Recall in Section \ref{sec:challenges}, our technique must overcome system challenges that are mainly attributed to noisy WiFi data; for example, device-specific activities do not reflect the user's actual physical state. To handle spurious events, we implement uncertainty quantification, which carries forward the prediction confidence in our ML model. Specifically, we define all prediction states:

\begin{equation}
    c_m = 
\begin{cases}
    0, & \text{if prediction mean within 95\% CI}\\
    1, & \text{otherwise}
\end{cases}
\end{equation}

Accordingly, we calculate the rate of estimation uncertainty instanced by the number of low confidence predictions within the sequence, $C = \{c_{1}, ..., c_{m}\}$. As will be discussed in Section \ref{sec:estimatingSleep}, the degree of tolerable uncertainty can be specified and only days with uncertainty less than that threshold are considered. In our case, our analysis found approximately 80\% of our sleep estimates within 5\% uncertainty rate, thus, regarding sequences with more than 5\% uncertainty rate as spurious sequences. 

\subsubsection{Cloud-based Implementation} 
Our models are built with Python, and the ML component is implemented using Keras with TensorFlow backend, \emph{scikit-learn} and \emph{forestci} libraries \cite{pedregosa2011scikit,abadi2016tensorflow,polimisconfidence}. The ML model is trained with a combination of a general dataset and 40\% of the user's data, thus semi-personalized. \system is built as a cloud-based web service, where the training of a semi-personalized model and testing of new data using random forest are completed in a matter of seconds. However, quantifying the uncertainty through the infinitesimal jackknife method is typically more expensive than obtaining the random forest.

\new{As our results will show in Section \ref{sec:evaluation}, the performance improvement using LSTM did not reach a significant difference compared to RF. While LSTM achieves a 3\% increase, this improvement is at the expense of precision and overall accuracy. Since RF is less computationally expensive, our solution is best suited to RF with infinitesimal jackknife method. Nonetheless, it should be noted that LSTM as another classification technique is applicable.} 

\section{User Study}
\label{sec:study}
To evaluate \system, we conducted an IRB-approved study, with all participants providing written informed consent before participating. Table \ref{tab:datasumm} summarizes our participants' data.

\subsection{Procedure and Participants}
Our primary user study involved 46 students in on-campus university dormitory housing. \new{Participants were recruited in batches between March-May 2021, and data collection was 4-weeks for each batch. The study took place during the COVID-19 pandemic. During this time, there were some restrictions regarding group size gatherings. As part of these restrictions, students attended their lessons online while living in their on-campus dormitories.} 

\new{In designing this study, our expert sleep researchers fully considered altering the procedure to its bare minimum. It should be noted that sleep studies with gold standard polysomnography (PSG) are typically conducted under conditions where the user's sleep is not interfered with. Having a diagnosed sleep disorder was an exclusion criterion during recruitment; however, there were no criteria for habitual sleep duration or times. Participants included both regular and irregular sleepers with varying sleep habits. Participants were required to own multiple devices, install a dedicated data logging app on their smartphone and actively maintain the diary log. With preserving user privacy in mind, log entries are limited to collecting only sleep and wake times, not reporting wakeful activities at night.}

\begin{table}[h!]
\centering
\caption{\new{Data summary of student participants in the primary user study.}}
\scalebox{.9}{
\begin{tabular}{l|l|l|l}\hline
\textbf{Users, N} & 46 (23 Male, 23 Female) & \textbf{Batch, n} & \makecell[l]{\#1 = 6, \#2 = 8, \#3 = 7, \\\#4: 6, \#5 = 6, \#6 = 13}\\\hline
\textbf{Age} & 20 - 28 years old, mean: 21.95, stdev.: 1.43 & \textbf{Year} & \makecell[l]{44 undergraduates, 2 graduates}\\\hline
\textbf{Study duration} & 4 weeks per user & \textbf{\makecell[l]{\% of time\\spent asleep}} & $\approx$ 12-20\% of the day per student \\\hline 
\textbf{Sleep summary} & \makecell[l]{Bedtime: 12:00 am - 5:30 am (mean: 1:47 am),\\Wake time: 5:30 am - 1:15 pm (mean: 8:29 am),\\Sleep duration: 195 - 660 mins (mean: 401 mins)} & \textbf{\makecell[l]{Residence}} & \makecell[l]{Self-contained student housing\\estate, with 30 blocks\\ segregated into 7 residences} \\\hline
\textbf{Oura baseline} & 14 - 24 days (mean: 17 days) & \textbf{\makecell[l]{Devices\\tracked}} & \makecell[l]{WiFi network activity data of\\ 1. smartphones, 2. laptop, 3. tablet} \\\hline
\end{tabular}}     
\label{tab:datasumm}
\end{table}

Upon consent, participants met at the start of the study to collect an Oura ring (gen 2) sleep wearable tracker for baseline sleep data; their participation was staggered in six batches. Participants also attended a one-time in-lab clinical assessment to assess their sleep health.  \new{Following standard sleep study procedures, participants fulfill a list of questionnaires, including the  Beck's Anxiety Inventory \cite{steer1997beck}, Beck's Depression Inventory \cite{steer2001mean}, and Berlin Questionnaire \cite{netzer1999using}. These questions were as sanity checks for us to be aware of their health states. All participants were in good standing in their mental health self-reports. Similarly, we identified no students with sleep apnea based on the Berlin questionnaire.} 

Throughout the study, participants were encouraged to utilize their on-campus residence WiFi frequently. We recorded all the MAC addresses of students' personal smartphones (Android or iOS), laptops, and tablets. \new{For students who owned iOS-based smartphones, they received a study-issued Android-based smartphone with a preinstalled data logging app.} We use the WiFi MAC addresses of their devices to extract the RTLS data of our participants. Note: by default, the MAC addresses in the RTLS data are hashed. Thus, we only identified individual users after explicitly providing us with their MAC addresses. At the end of the study, all study-issued devices were returned, and each participant received compensation of up to USD 18.12 weekly in cash if they fulfilled all study requirements. These requirements include: (a) regularly wearing the Oura ring and logging their sleep, (b) installing the required data logging app on their smartphone, and (c) using the campus WiFi while in their dorms.

We extracted WiFi network activity data for all user-owned devices for the days when their Oura baseline data and self-reports were correct and available. Note: the distribution of sleep and awake states for each user is highly imbalanced as users spent approximately 20\% of the day sleeping.

\begin{table}[h!]
\caption{\new{Data summary of home participants in the supplementary user study.}}
\scalebox{.9}{
\begin{tabular}{c|c|C{4.5cm}|C{4cm}}\hline
& $HomeUser_{1}$ & $HomeUser_{2}$ & $HomeUser_{3}$ \\ \hline
\textbf{User} & Male & Female & Male \\\hline
\textbf{Residence} & Single-family house & \multicolumn{2}{c}{Apartment} \\ \hline
\textbf{Household} & Family with children & \multicolumn{2}{c}{Couple no children}\\ \hline
\textbf{Study duration} & 1 month & \multicolumn{2}{c}{1 week} \\ \hline
\textbf{Devices tracked} & \makecell[l]{smartphone, laptop} & \multicolumn{2}{c}{\makecell[l]{smartphone (personal), laptop (personal), smartTV, smart-speaker}} \\ \hline
\textbf{Sleep baseline} & Diary logs & \multicolumn{2}{c}{Fitbit + Diary logs + lights off log} \\ \hline
\textbf{Sleep summary} & \makecell[l]{Bedtime: 12:00 am\\Waketime:\\ 6:00 am (weekday),\\ 7:00 am (weekend)\\Sleep duration: 360 mins} & \makecell[l]{Bedtime: 11:20 pm -12:45 am\\ (mean: 11:46 pm)\\Waketime:\\ 5:30 - 8:00 am\\(mean: 6:39 am)\\Sleep duration: 290 - 480 mins \\(mean: 376 mins)} & \makecell[l]{Bedtime: 11:15 pm - 12:30 am\\(mean: 11:37 pm)\\Waketime:\\ 5:40 - 8:00 am\\(mean: 6:45 am)\\Sleep duration: 310 - 450 mins\\ (mean: 372 mins)} \\ \hline
\textbf{Lights off} & - & \multicolumn{2}{c}{Weekday: 11:00pm, Weekend: 11:30 pm} \\ \hline
\textbf{\makecell[l]{\% of time\\spent sleeping}} & 25\% of the day & 26\% of the day & 25\% of the day \\ \hline
\end{tabular} }  
\label{tab:homeusers}
\end{table}

\new{To demonstrate the applicability of our approach among other populations, our study expands to include non-student users living in different private home settings, albeit on a smaller scale. The sleep baseline data for this set of participants is based on their preferred choice of a diary log and/or (their personal) Fitbit device. These logging modalities have also proven reliable for sleep monitoring purposes \cite{de2019wearable}. Fitbit provides similar functionalities and data files as the Oura ring.} Unlike student users, home participants directly provided us with their WiFi event logs in a .csv file. \new{Additionally, $HomeUser_{2}$ provided the logs of their share smart home devices. We summarize the demographic profiles of our home participants in Table \ref{tab:homeusers}.}

\subsection{Sleep Baseline}
\label{sec:expertModifiedData}
Each student participant wore an Oura sleep sensing ring as baseline sleep data to compare against \system. Students were asked to select an Oura ring size that was most comfortable. \new{Our home participants used their personal Fitbit, also equipped with similar features.} These wearables estimate sleep using heart rate, HRV (heart rate variability), body temperature, and movement via infrared photoplethysmography (PPG), temperature sensor, and a 3-D accelerometer signals \cite{altini2021promise}. Participants were instructed to wear their devices (both during the day and night) and sync the data to the Oura/Fitbit app daily. From this cloud data, daily sleep measures such as bedtime, wake time, time-in-bed (TIB), wake after sleep onset (WASO) as well as thirty-second epoch by epoch sleep stages data (wake, light, deep, and REM) were extracted using the respective cloud API. Note: we only use the wearables as a baseline and not as ground truth for medically approved sleep studies requiring highly invasive and expensive polysomnography. 

\new{To ensure the baseline sleep data was correct and reliable, we used the sleep summary file containing only duration and time information and the hypnogram file containing wake and sleep stages every 30 seconds. Additionally, we used self-reports to check the alignment of sleep and activity data with self-reports.}

\subsubsection*{Practical Challenges}
\new{First, insufficient battery power and missed opportunities to wear the device often led to inaccurate sleep estimation among our students -- this is a common challenge faced in many sleep studies \cite{de2019sleep,altevogt2006sleep,massar2021trait}. Second, the Oura ring was not equipped with a nap detection feature at the time of the study. A common but small observation among our students was that some morning naps were regarded as a continuation of nocturnal sleep (mismatch between sleep summary and hypnogram). Due to discrepancies and inadequate ground truth to validate these occurrences, we made the operational decision to exclude training our model for nap-like sleep. Approximately 10\% of all data points exhibited this pattern. Hence, we excluded data if there is a mismatch in sleep timings between these three resources. Given a month's participation per student, the data exclusion rate from inaccurate reportings ranged between 7\% to the highest of 50\%. We excluded no data from home users.}

\subsection{Ethical Consideration} 
A key consideration in this work is preserving users' privacy while still being functional. In practice, WiFi data can reveal many aspects of users' private information. \system is designed to keep user privacy in mind by collecting coarse-grain network activity. As such, \system only uses network data without knowing the corresponding user activity for accessing WiFi or location information derived from approximating AP locations. In addition, \system will only provide user-specific sleep predictions if the user explicitly provides the MAC addresses of their devices. It is important to note that network activity data of student devices were directly collected from the campus infrastructure and bounded by the computing agreements agreed to by each user when they received their WiFi credentials. Without these MAC addresses to user device mappings, \system can still produce sleep results at aggregate levels but without attributing results to users. Aggregated results remain helpful as they can provide a good health status overview of an entire home or residential dormitory without needing to identify individual users.
\section{Evaluating \system}
\label{sec:evaluation}
We evaluate the performance of \system using the dataset collected from our primary user study. \new{We conduct our model evaluation through a leave-one(user)-out cross-validation. As will be explained later, we build a semi-personalized model where a model for each user is trained using 40\% of their data. Using all but the user's data as part of the training set, we repeat this process five times, where each training time includes a different set of random samples (of the test users) to build the semi-personalized model. We present the accuracy, recall, precision, and F1 scores (weighted average of precision and recall).}

\subsection{Efficacy of ML Models}
\label{sec:resultML}
Our first evaluation compares the efficacy of different algorithms at predicting sleep and awake states in 24 hours. Building on prior findings \cite{mammen2021wisleep}, we contrast the performance of Naive Bayes (NB), Random Forest (RF), and Gradient Boosting (XGBoost) sampling features at every 15 minutes interval. As shown in Table \ref{tab:modelML}, a generalized model using random forest yields the highest accuracy of 79.30\% in classifying sleep states. 

\begin{table}[ht]
\centering
\caption{Model efficacy with different ML algorithms, and random forest yielding best results.}
\scalebox{.9}{
\begin{tabular}{l|c|c|c|c|c}\hline
\textbf{Algorithm} & \textbf{Accuracy (Acc)} & \textbf{Precision (Prec)} & \textbf{Recall (Rec)} & \textbf{F1} & \textbf{p} \\ \hline
NB & 0.538	& 0.394 & 0.833 & 0.521 & p<.01 \\
XGBoost & 0.775 & 0.719 & 0.425 & 0.498 & p<.01 \\
RF & 0.793 & 0.693 & 0.582 & 0.624 & - \\
RF (tuned) & 0.798 & 0.713 & 0.574 & 0.625 & p<.01\\
\hline
\end{tabular} }     
\label{tab:modelML}
\end{table}

\begin{table}[ht]
\begin{minipage}[b]{0.45\linewidth}
\centering
\caption{Training data for RF model personalization.}
\scalebox{.9}{
\begin{tabular}{c|c|c|c|c|c}\hline
\textbf{Train} & \textbf{Acc} & \textbf{Prec} & \textbf{Rec} & \textbf{F1} & \textbf{p} \\ \hline
General & 0.798 & 0.713 & 0.574 & 0.625 & -\\
+ 10\% user & 0.825 & 0.751 & 0.632 & 0.681 & p<.01 \\
+ 20\% user & 0.835 & 0.761 & 0.666 & 0.706 & p<.01 \\
+ 30\% user & 0.844 & 0.771 & 0.682 & 0.720 & p<.05 \\ 
+ 40\% user & 0.851 & 0.788 & 0.697 & 0.736 & p<.05 \\
+ 50\% user & 0.858 & 0.795 & 0.714 & 0.750 & p>.1 \\
\hline
\end{tabular} }    
\label{tab:modelMLper}
\end{minipage}
\hspace{0.5cm}
\begin{minipage}[b]{0.48\linewidth}
\centering
\caption{Sample frequency for RF semi-personalized model.}
\scalebox{.9}{
\begin{tabular}{c|c|c|c|c|c}\hline
\textbf{Frequency} & \textbf{Acc} & \textbf{Prec} & \textbf{Rec} & \textbf{F1} & \textbf{p} \\ \hline
15 minutes & 0.851 & 0.788 & 0.697 & 0.736 & - \\
30 minutes & 0.815 & 0.731 & 0.625 & 0.666 & p<.01 \\
10 minutes & 0.872 & 0.817 & 0.746 & 0.777 & p<.05 \\
5 minutes & 0.906 & 0.860 & 0.824 & 0.840 & p<.01 \\
\textbf{1 minute } & \textbf{0.939} & \textbf{0.896} & \textbf{0.905} & \textbf{0.900} & p<.01 \\ 
45 seconds & 0.92 & 0.876 & 0.887 & 0.880 & p>.1 \\
30 seconds & 0.926 & 0.875 & 0.888 & 0.878 & p>.1\\
\hline
\end{tabular} }     
\label{tab:cutoffFreq}
\end{minipage}
\end{table}

\begin{figure}[h!]
\centering
\vspace{-0.2in}
\includegraphics[width=.8\columnwidth]{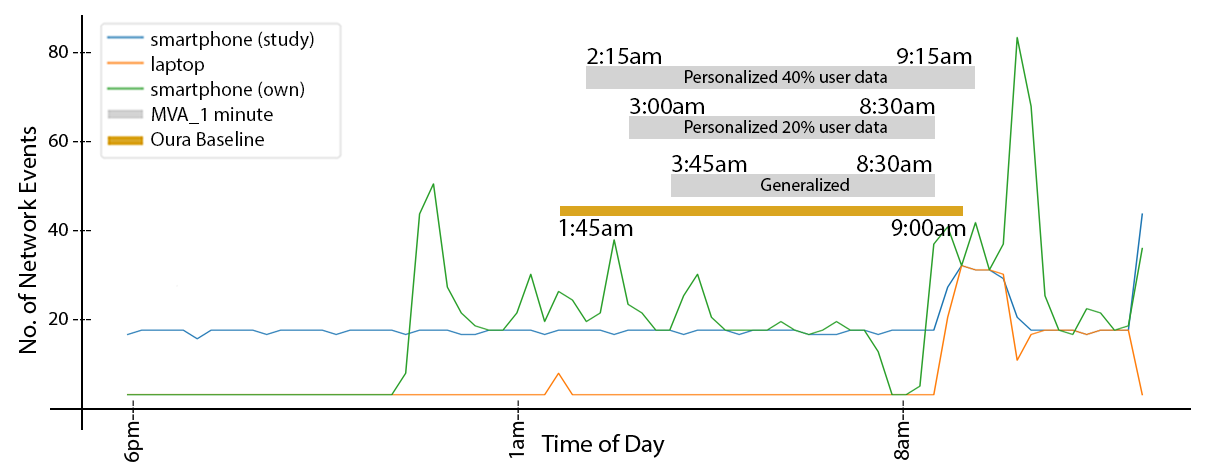}
\caption{Personalization improves sleep prediction.}
\label{fig:perGenUser}
\end{figure}

\subsubsection{Model Personalization}
\label{sec:modpersonalize}
Next, we compare the difference between using generalized and personalized models where the personalized models were trained using a portion of the test user's data. Overall, the personalized models improve model performance by a small but significant percentage. As shown in Table \ref{tab:modelMLper}, using 10\% training data to build a per-user model improves model accuracy by $\approx$ 3\% compared to a generalized model (p<.01). Recursively adding 10\% of training data slightly improves the model accuracy by $\approx$ 1\%. However, the performance improvements did not achieve a statistical difference beyond 40\% training data.

Model personalization also helps to improve recall -- the probability that a sleep state is correctly predicted. Figure \ref{fig:perGenUser} illustrates the prediction outcomes for user P23 over a night. With 40\% training data, recall and overall accuracy improved by 3\% and 90\% compared to a generalized approach (88\%). 

\subsubsection{Sampling Frequency} \label{sec:frequencysensitivity} Another key input to \system is the WiFi network activity data sampling rate. In Section \ref{sec:rtls}, we explained that the RTLS data frequency ranges from every 5 seconds to several minutes depending on device use. Guided by prior work~\cite{mammen2021wisleep}, we investigate the performance improvements at different WiFi data sampling rates of 30 seconds to 30 minutes. Our analysis, shown in Table \ref{tab:cutoffFreq}, found that reducing 15 minutes sampling frequency up to 1 minute significantly improves the overall model performance and recall by $\approx$ 10\%. Further decreasing the sampling rate to 45 or 30 seconds has a slight statistically insignificant performance reduction. Thus, we used a WiFi sampling rate of 1 minute. 

\subsubsection{Model Tuning} 
\label{sec:modeltune}
The inner workings of our chosen classification algorithm random forest use multiple decision trees for decision-making. Fine-tuning our model yields $\approx$ 94\% accuracy, although it is not statistically significant compared to the default model. Several essential parameters for tree-based classification algorithms are the number of trees (n\_estimator) and the number of data points placed in a node before it is split (min\_samples\_split) to improve overfitting. We iteratively optimized the hyperparameters of our model using grid search to reduce bias in calculating the variance estimates \emph{n\_estimator=200}, and \emph{min\_samples\_split=5}. 

\subsubsection{Efficacy of DL Model}
\label{sec:resultDL}
\new{Our implementation of LSTM begins with a single layer and progressively adding layers to achieve a robust model. Similarly, we used the ten features described in Section \ref{sec:features}, sampled at 1-minute interval, as the input to the model, and incrementally increased the adjacent samples (i.e., timestep) from 5 to 30 minutes intervals to predict each instance. For example, we set the ``lookback period'' to every 10 minutes of data in the past to predict the upcoming sample.}

\begin{table}[h]
\centering
\caption[table]{\new{Comparing \system with LSTM network.}}
\scalebox{.9}{
\begin{tabular}{l|c|c|c|c|c}\hline
\textbf{Model} & \textbf{Accuracy (Acc)} & \textbf{Precision (Prec)} & \textbf{Recall (Rec)} & \textbf{F1} & \textbf{p} \\ \hline
LSTM-5T & 0.859 & 0.841 & 0.602 & 0.702 & -\\%
LSTM-10T & 0.832 & 0.763 & 0.676 & 0.717 & p<.01\\%
LSTM-15T & 0.858 & 0.791 & 0.661 & 0.720 & p<.01\\%
\textbf{LSTM-30T} & \textbf{0.928} & \textbf{0.879} & \textbf{0.883} & \textbf{0.881} & p<.01\\ \hline
\textbf{LSTM-2} & \textbf{0.930} & \textbf{0.863} & \textbf{0.913} & \textbf{0.887} & p>.1\\%
LSTM-3 & 0.921 & 0.866 & 0.876 & 0.872 & p>.1\\%
\hline
\end{tabular} }     
\label{tab:modelNN}
\end{table}

\new{Table \ref{tab:modelNN} summarizes the model performance, first, on a single-layer LSTM model with ten hidden layers, 32 samples in each mini-batch, and an epoch size of 5, with varying timesteps. We set the dropout and recurrent dropout rates at 20\% as a starting point. Where recall is prioritized, the model yields the best results of 88.3\% recall and 92.8\% accuracy at 30 minutes time step.} 

\begin{figure}[h!]
\begin{minipage}[b]{0.37\linewidth}
\includegraphics[width=\columnwidth]{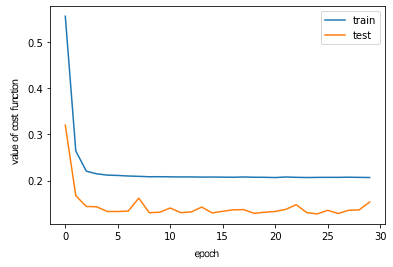}
\caption{Value of cost function decreases with epoch size for P10.}
\label{fig:LSTM1}
\end{minipage}
\hspace{0.03cm}
\begin{minipage}[b]{0.61\linewidth}
\includegraphics[width=\columnwidth]{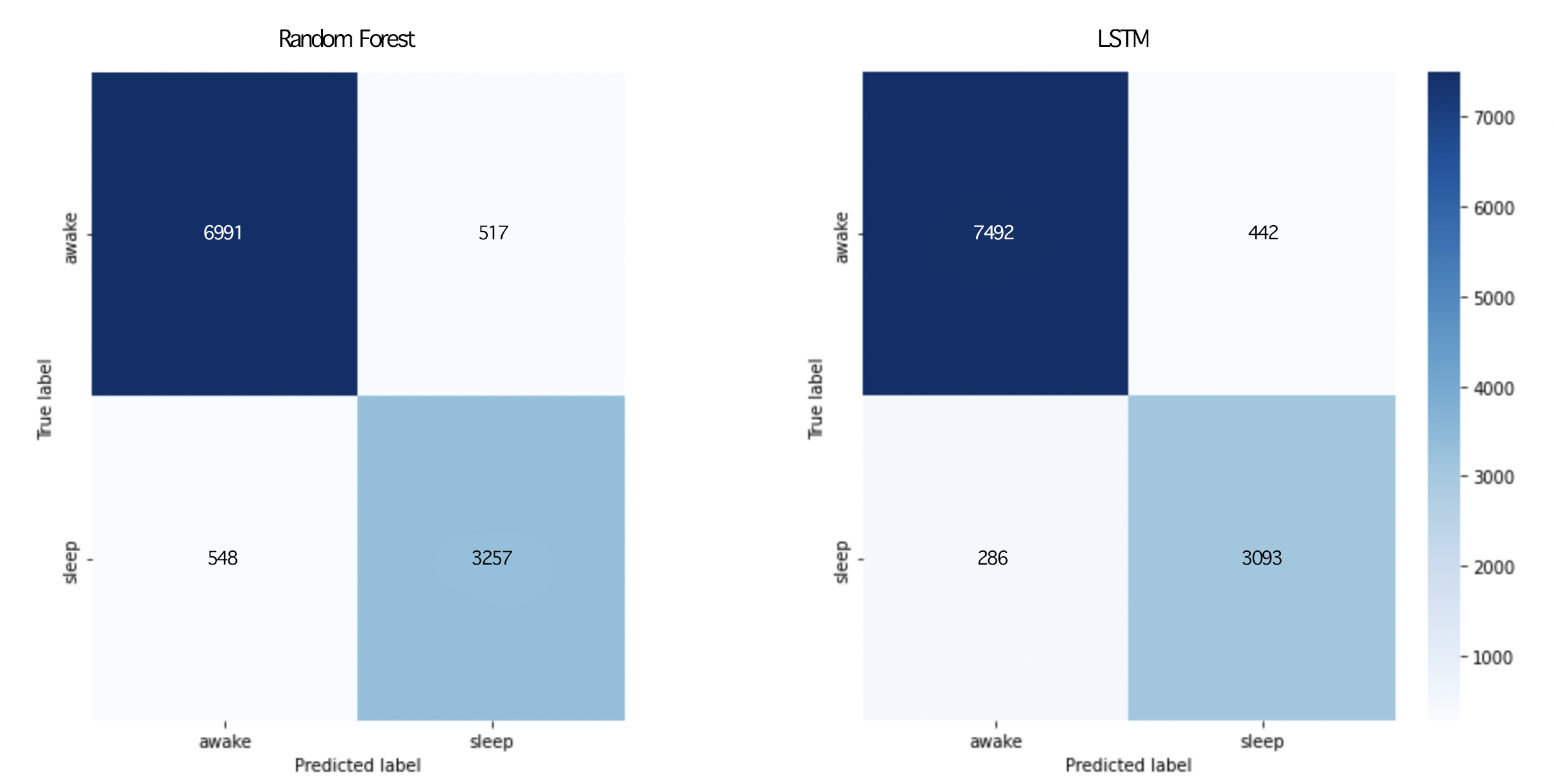}
\caption{Confusion matrix comparison between RF and LSTM for P10.}
\label{fig:LSTM2}
\end{minipage}
\end{figure}

\new{The subsequent addition of an LSTM layer (i.e., LSTM-2) and fine-tuning improve recall. However, we quickly observed declining performance with a third LSTM layer. As the LSTM model is slower to train, we implement a grid search for LSTM-2, attending to the following hyperparameters: \emph{lstm} and \emph{recurrent} dropout rates, \emph{epoch}, and \emph{batch size}. With the optimal hyperparameters at \emph{lstm\_dropout=0.01}, \emph{recurrent\_dropout=0.2}, \emph{batch\_size=64}, and \emph{epoch=5}, a two-layer LSTM model yields best performance at 91.3\% recall, increasing recall by 3\% compared to a single-layer LSTM. However, this difference is not significant. Figure \ref{fig:LSTM1} charts the value of cost function, that is, the error between predicted values and true values, over the entire training set for an example user, P10. Contrasting model performance with RF, we plot the confusion matrix for the same user in Figure \ref{fig:LSTM2}.}

\new{While promising, the average performance improvement using LSTM was only on the recall measure and is not significantly different from the RF algorithm. This minimal improvement did not justify building a complex model that is also computationally expensive compared to an RF model. Many applications utilizing LSTM networks have proven successful on large amounts of training data (e.g., months and years) to detect patterns of residential and student life behaviors \cite{bovornkeeratiroj2021vpeak,umematsu2019improving}. Until our application has collected much more data to build classifiers that compare the upcoming sleep duration for the day to the sleep histories of the same days in previous weeks/months, we reckon using the RF algorithm at this stage for our system to function adequately. We discuss our continuing effort for sleep prediction using LSTM  in Section \ref{sec:discussion}. Note that our results are based on the RF algorithm in the experiments moving forward. }

\subsection{Benefit of Using Multiple Devices}
\label{sec:featureanalysis}
A key hypothesis of this work is that using data collected from multiple devices owned by a user will lead to better sleep prediction than using just a single device. To test this hypothesis, we compare the accuracy of using multiple devices versus a single preferred primary device. A preferred device, in this case, is a smartphone, utilized most frequently \cite{van2015modeling}.

\begin{table}[h!]
\centering
\caption{Multi-device connection to WiFi per day.}
\scalebox{.9}{
\begin{tabular}{l|c|c}\hline
\textbf{Device Type} & \textbf{No. of Users with Device (\%)} & \textbf{\% WiFi activity per day}\\\hline
Smartphone & 46 (100\%)  & 47.30\% \\
Smartphone (study) & 13 (28\%) & 27.10\% \\
Laptop & 39 (85\%) & 35.19\%  \\
Tablet & 10 (22\%) & 33.26\% \\
\hline
\end{tabular}}     
\label{tab:primaryDevice}
\end{table}

Table \ref{tab:primaryDevice} summarizes the percentage of time different user-owned devices were connected to the WiFi network. In particular,  smartphones were owned by 100\% of our users and connected to the monitored WiFi network $\approx$ 50\% of the time. Note: the monitored network was only in the dorms; users were probably elsewhere the rest of the time. Laptops (owned by 85\% of our participants) were connected $\approx$ 35\% of the time.

\subsubsection{Performance of Multi-Device Features} 
\begin{figure}[h!]
\centering
\includegraphics[width=.65\columnwidth]{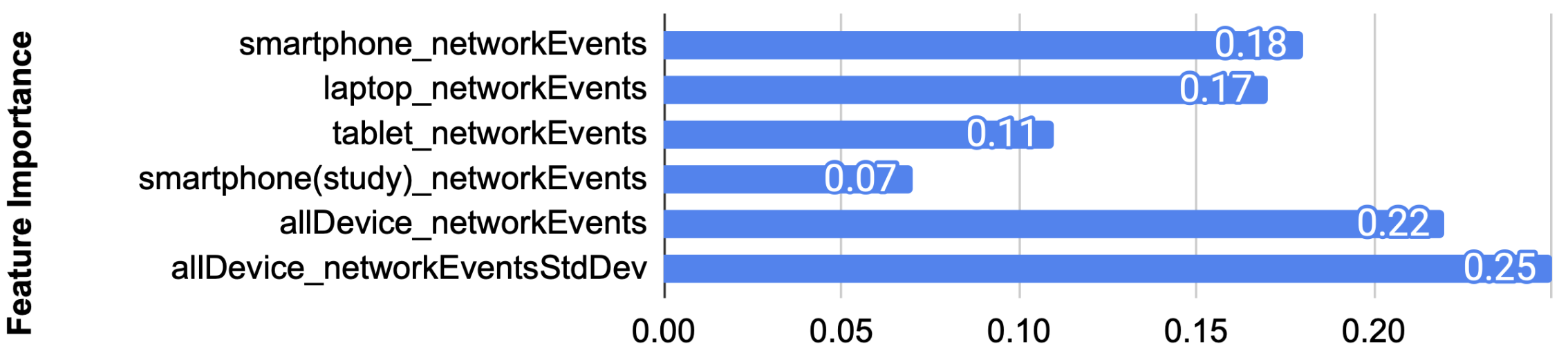}
\caption{Feature importance.}
\label{fig:featureImportance}
\end{figure}

Figure \ref{fig:featureImportance} shows the importance of all features based on our model's fitted trees. The F-score for feature importance is measured in terms of weight, that is, the number of times the feature is used in a tree. Intuitively, a high F-score (1 being the max) reflects how important the feature is independent of other features. This analysis revealed that using the sum of network events across all user-owned devices had the highest F-score compared to using events only from smartphones or tablets.

\begin{table}[h!]
\centering
\caption{Performance with different device features.}
\scalebox{.9}{
\begin{tabular}{l|c|c|c|c|c}\hline
\textbf{Feature} & \textbf{Accuracy} & \textbf{Precision} & \textbf{Recall} & \textbf{F1} & \textbf{p-value} \\ \hline
Smartphone & 0.893 & 0.860 & 0.781 & 0.812 & - \\
Smartphone + 1 secondary device & 0.911 & 0.851 & 0.863 & 0.855 & p<.01\\
Smartphone + multi devices & 0.939 & 0.896 & 0.905 & 0.900 & p<.01 \\
\hline
\end{tabular} }     
\label{tab:featureComparison}
\end{table}

Table \ref{tab:featureComparison} compares the model performance with different device features. Using just smartphones yields a 78.10\% recall and 89.30\% accuracy. However, adding more than two user device features increases recall by more than 10\% and overall accuracy by 4\% (p<.01). The results strongly suggest that using multiple devices will improve \system's accuracy and recall. 

\subsection{Efficacy of Sleep Estimation}
\label{sec:estimatingSleep}
With the classification component determined, \system now has to interpret the classification results. In particular, given a continuous prediction of sleep and awake states, \system must now accurately predict the start and end of a user's sleep period ($T_{sleep}$, $T_{wake}$) and the sleep duration.

\subsubsection{Choice of Smoothing Technique}
\begin{figure}[h!]
\centering
\includegraphics[width=.95\columnwidth]{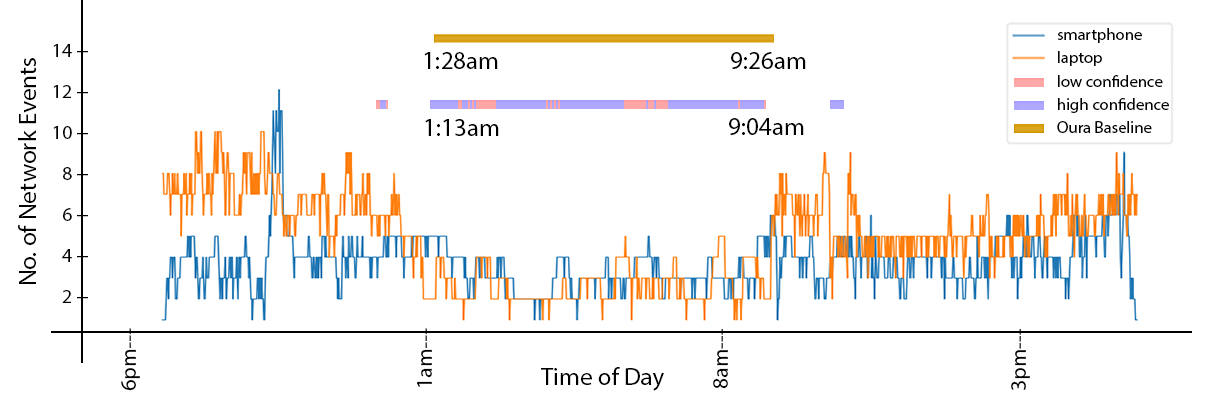}
\vspace{-0.1in}
\caption{\new{Predicted states with confidence level for P10.}}
\label{fig:smoothingProblem}
\end{figure}

\new{Unfortunately, coupling the confidence interval for each predicted outcome does not provide a straightforward solution to determining a user's sleep duration. For example, a typical representation is depicted in Figure \ref{fig:smoothingProblem} where we observed most sleep states with low confidence spuriously occurring between 1:00 am, and 7:00 am. Ignoring the first and last occurrence of low confidence predictions does not directly translate to estimating the start and end of a sleep period. For these reasons, we described two techniques to estimate sleep timing based on the most extended sleep period in Section \ref{sec:estmodel}: predicting with moving averages (MVA) and smoothed aggregation (AGG). Instead, confidence values will be used to quantify the uncertainty rate of an entire sleep duration prediction once it is estimated.}

MVA determines a \emph{sleep} or \emph{awake} state by gradually moving the window, defined by the threshold interval, over the predicted outcomes in single increments. Then, we determine the most extended sleep period from the revised outcomes. In AGG, we total the sum of every few sleep states, defined by the threshold interval, and smooth out the representation by applying a Savitzky-Golay (SG) filter \cite{press1990savitzky}. 

Table \ref{tab:finalerrors} tabulates the mean error in users' sleep period with varying window sizes. The errors of $T_{sleep}$ and $T_{wake}$ are calculated as the difference in time between our sleep estimates and the users' Oura ring sleep baseline reference. We observed that using MVA with a rolling window size of 5 to 20 minutes hovered between 39 to 41 minutes in estimating $T_{sleep}$ $T_{wake}$ errors. However, with AGG, errors fluctuate in a more extensive range of 42 to 95 minutes. With \system achieving the lowest mean errors in single increments of 5-minutes (regardless of method), we adopt this window size for the MVA estimation method moving forward.

\begin{table} [!h]
\caption{Comparison of errors in minutes using two estimation methods on predictions with varying window size (W).}
\scalebox{.9}{
\centering
  \begin{tabular}{l|c|c|c|c|c|c|c|c|c}\hline
    \multirow{2}{*}{\textbf{Method}} &
      \multicolumn{2}{c|}{\textbf{W=5}} &
      \multicolumn{2}{c|}{\textbf{W=10}} & 
      \multicolumn{2}{c|}{\textbf{W=15}} &
      \multicolumn{2}{c|}{\textbf{W=20}} & \textbf{p-value} \\
      & $T_{sleep}$ & $T_{wake}$ & $T_{sleep}$ & $T_{wake}$ & $T_{sleep}$ & $T_{wake}$ & $T_{sleep}$ & $T_{wake}$ & \\ \hline
      MVA &  39 & 37 & 39 & 37 & 40 & 37 & 41 &	37 & p<.01 \\\hline
      AGG &  47 & 42 & 65 & 56 & 80 & 66 & 95 & 78 & - \\\hline
  \end{tabular}}
  \label{tab:finalerrors}
\end{table}

\subsubsection{Improving Performance with Uncertainty Rate} 
These errors thus far include the presence of sleep estimates with significant uncertainty. Recall in Section \ref{sec:estmodel} that the uncertainty rate is instanced by the number of states predicted with low confidence. We consider only sleep estimates with less than 5\% uncertainty to improve the system's performance, making up roughly 80\% of our prediction samples. 

\begin{figure}[h!]
\begin{minipage}[b]{0.65\linewidth}
\includegraphics[width=\columnwidth]{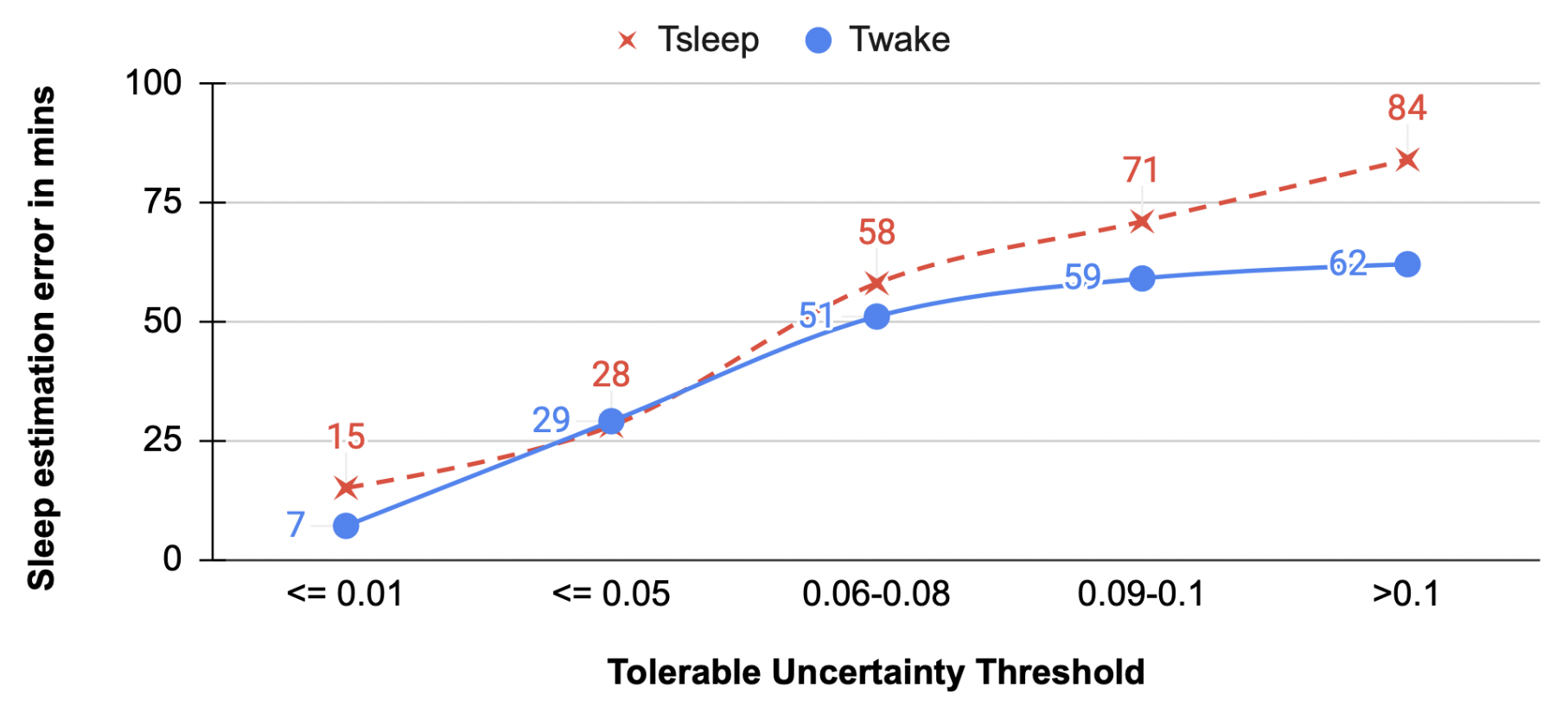}
\vspace{-0.2in}
\caption{Sleep estimation error in minutes at varying uncertainty rates.}
\label{fig:estimatedTimes}
\end{minipage}
\hspace{0.03cm}
\begin{minipage}[b]{0.33\linewidth}
\includegraphics[width=\columnwidth]{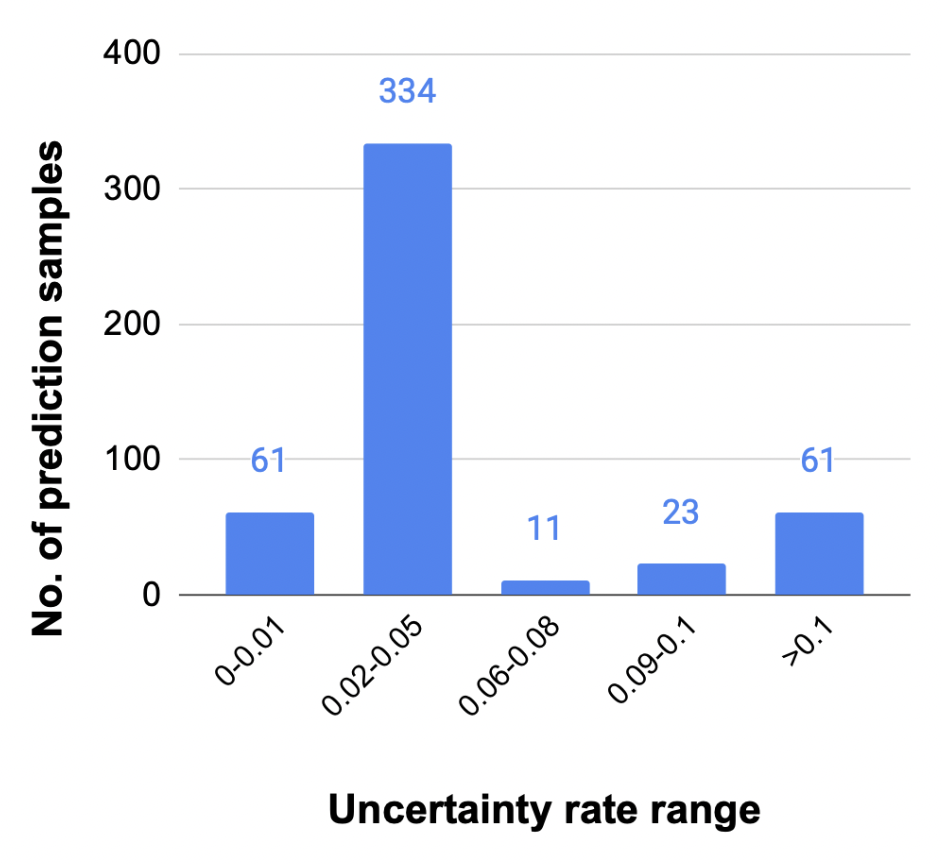}
\vspace{-0.2in}
\caption{Prediction sample distribution.}
\label{fig:predDistribution}
\end{minipage}
\end{figure}

Figure \ref{fig:estimatedTimes} charts our estimation errors with varying uncertainty rates. Sleep estimations made within 1\% uncertainty yielded the lowest errors of 15 minutes for sleep time and 7 minutes for wake times. Estimates made within 5\% uncertainty achieved an average error of 28 minutes sleep time and 29 minutes wake time. As the tolerance threshold increases, the technique ``filters'' out more days with an uncertainty rate higher than the threshold and provides higher accuracy. This mechanism results in a tradeoff where the number of days made by a prediction depends on the tolerance. The higher the tolerance, the more the cost of higher errors. 

\new{In keeping the threshold within 5\% uncertainty, \system retains 80.6\% of its prediction outcomes. As per Figure \ref{fig:predDistribution}, most of our predictions are made within 0.02-0.05 uncertainty rate range, thus, excluding 95 of 490 outcomes. Restricting the threshold to 1\% retains 12\% of these outcomes.}

\subsection{Summary of Findings} In its present form, \system predicts users' sleep states by employing a fine-tuned semi-personalized (using 40\% of a user's data) random forest classification model and infinitesimal jackknife variance estimation method to produce confidence intervals. We collected, on average, 17 days of data per user in our data collection study -- thus, six days' worth of user data would be used for training. Our approach samples WiFi-generated features from multiple user devices every minute. In practice, \system does not need to sample WiFi data for the entire day consistently; instead, it simply needs to sample during reasonable sleeping periods every day (i.e., at the very least from 8 pm to the following morning on weekdays with afternoons added on weekends). We provide quantitative evidence to strongly suggest that using multiple devices will improve \system's accuracy and recall. Finally, using MVA-5 with a rolling window size of 5 minutes yielded the best results in estimating the longest sleep period. The sleep errors range between 15-28 minutes and wake errors range between 7-29 minutes within a 5\% uncertainty rate.

\section{Comparative Analysis} 
\label{sec:errorAnalysis} 
\new{This section focuses on the sleep estimates produced by \system with different devices, including the Oura ring, chosen as a dedicated sleep tracker in our primary study. We draw comparisons to understand its performance with varying sleep behaviors and tested our system in a home setting and prior work utilizing AI techniques.}

\subsection{\system and Smart Devices}

\subsubsection{WiFi Data from Multiple Personal Devices}
Table \ref{tab:summaryEstimationError} breaks down our sleep estimation errors using an MVA-5 method by summarizing the complete statistics with WiFi data from varying personal devices. We observed that the errors occurring most often were 5 minutes for bedtime and 1 minute for wake time. However, the mean values and standard deviations tend to increase with lesser user devices, reiterating the value of adding more devices to achieve accuracy in sleep monitoring.

\begin{table}[h!]
\centering
\caption{Summary statistics of sleep and wake estimation errors with varying devices in minutes. }
\scalebox{.9}{
\begin{tabular}{r|C{2cm}|C{2cm}|C{2cm}|C{2cm}|C{2cm}|C{2cm}}\hline
\textbf{} & \multicolumn{2}{c|}{\textbf{Smartphone + multi devices}} & \multicolumn{2}{c|}{\textbf{Smartphone + 1 device}} & \multicolumn{2}{c}{\textbf{Smartphone only}}\\
 & $T_{sleep}$ & $T_{Wake}$ & $T_{sleep}$ & $T_{wake}$ & $T_{sleep}$ & $T_{wake}$ \\ \hline
\textbf{Median} & 7 & 7 & 18 & 15 & 21 & 27\\
\textbf{Mean} & 17 & 21 & 31 & 32 & 42 & 37\\
\textbf{Max} & 182 & 177 & 211 & 403 & 301 & 253\\
\textbf{Min} & 0 & 0 & 0 & 0 & 0 & 0\\
\textbf{Mode} & 5 & 0 & 5 & 1 & 5 & 1\\
\textbf{Stdev.} & 27 & 32 & 35 & 47 & 49 & 43\\
\textbf{Q1, Q3} & 5,16 & 2,26  & 6,43 & 4,42 & 8,58 & 3,51\\
\textbf{UIF, UOF} & 33, 51 & 63, 100 & 99, 156 & 99, 156 & 134, 210 & 123,195\\
\hline
\end{tabular} }     
\label{tab:summaryEstimationError}
\end{table}

\begin{figure}[h!]
\begin{minipage}[b]{0.49\linewidth}
\includegraphics[width=\columnwidth]{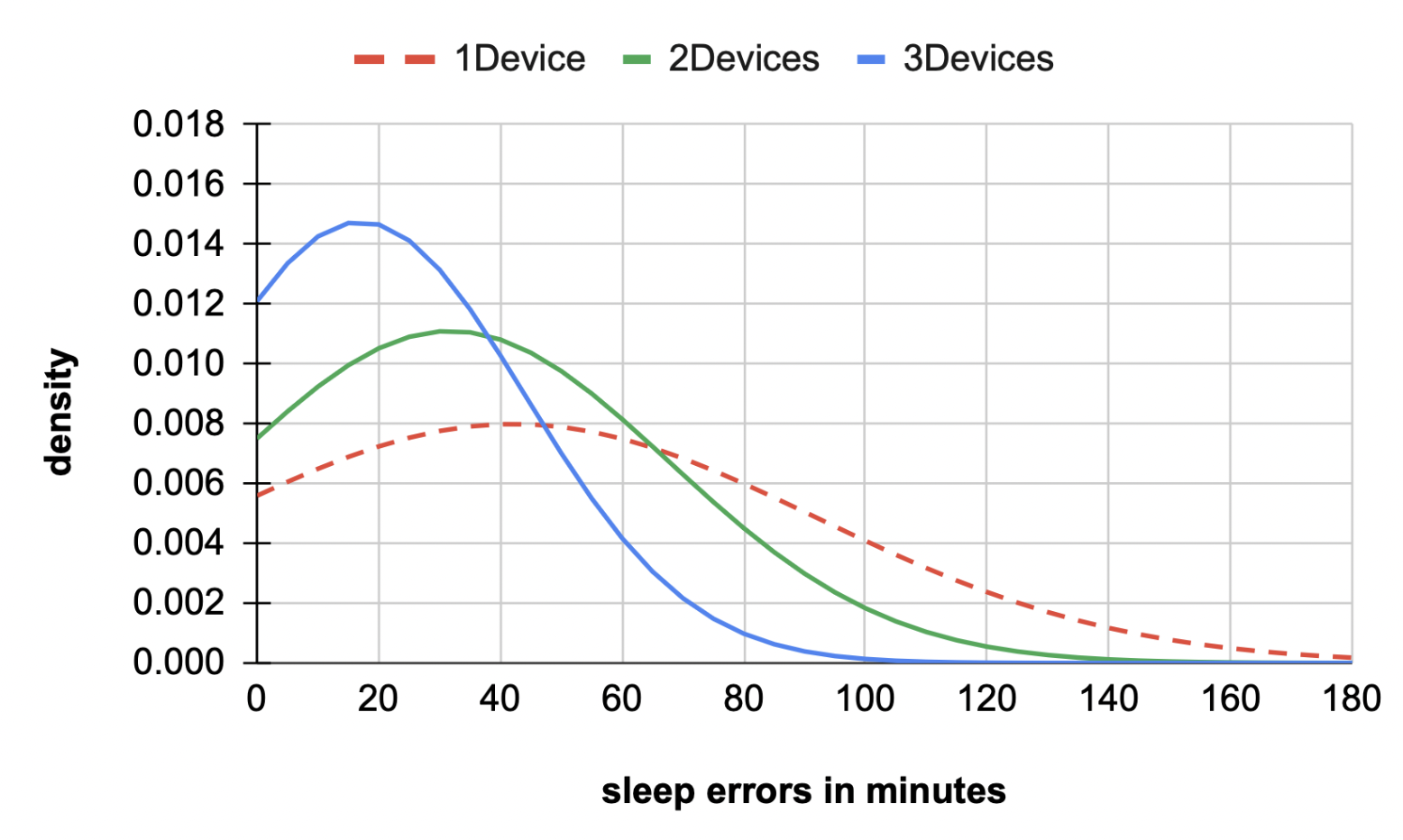}
\caption{PDF of sleep error with varying user devices.}
\label{fig:sleepdurationPDF}
\end{minipage}
\hspace{0.05cm}
\begin{minipage}[b]{0.49\linewidth}
\includegraphics[width=\columnwidth]{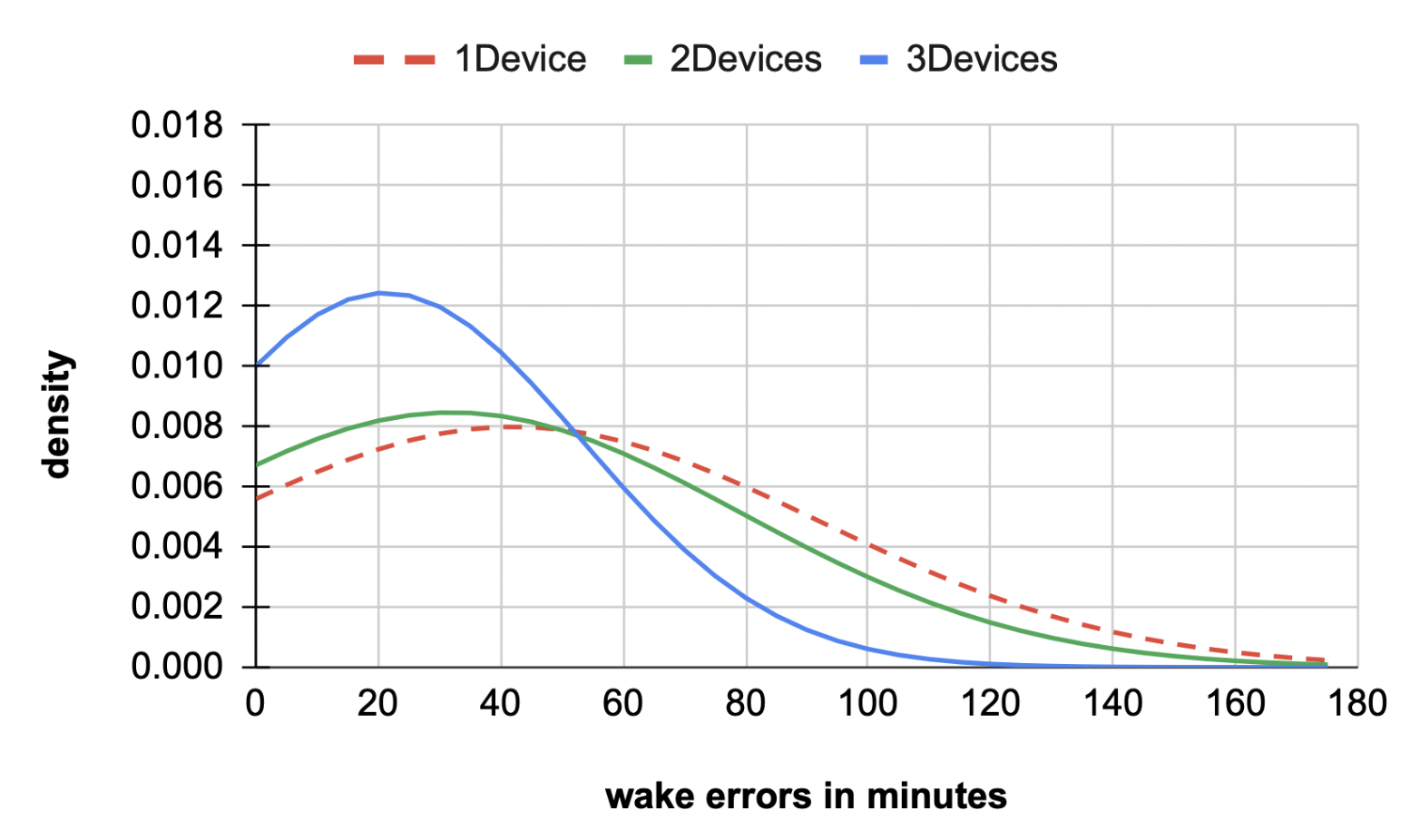}
\caption{PDF of wake error with varying user devices.}
\label{fig:wakedurationPDF}
\end{minipage}
\end{figure}

A probability density function view of the estimation errors in Figures \ref{fig:sleepdurationPDF} and \ref{fig:wakedurationPDF} also suggest that the outliers were a small fraction of the predictions with more devices used for monitoring. Sleep errors were reduced significantly with more than two secondary devices compared to using one smartphone (p<.01) and two devices (p<.01). Wake errors did not significantly improve by adding a secondary device compared to using a primary smartphone (p>.1) but significantly improved with multiple secondary devices (p<.01).

\subsubsection{Contrasting the Oura Ring}
\begin{figure}[h!]
\begin{minipage}{0.48\textwidth}
\captionof{table}{Descriptive statistics for Oura ring and \system.}
\scalebox{.9}{
\begin{tabular}{r|C{2.5cm}|C{2.5cm}}\hline
& $Oura_{duration}$ & $SleepMore_{duration}$ \\ \hline
\textbf{Median} & 404 & 430 \\
\textbf{Mean} & 400 & 426 \\
\textbf{Max} & 680 & 641 \\
\textbf{Min} & 240 & 210 \\
\textbf{Mode} & 428 & 428 \\
\textbf{Stdev.} & 75 & 67 \\
\textbf{Q1, Q3} & 358,448 & 389,471 \\
\textbf{UIF, UOF} & 582,717 & 594,718 \\
\hline
\end{tabular} }  
\label{tab:sleepdurationCompare}
\end{minipage}
\begin{minipage}{0.48\textwidth}
\centering
\includegraphics[width=1\columnwidth]{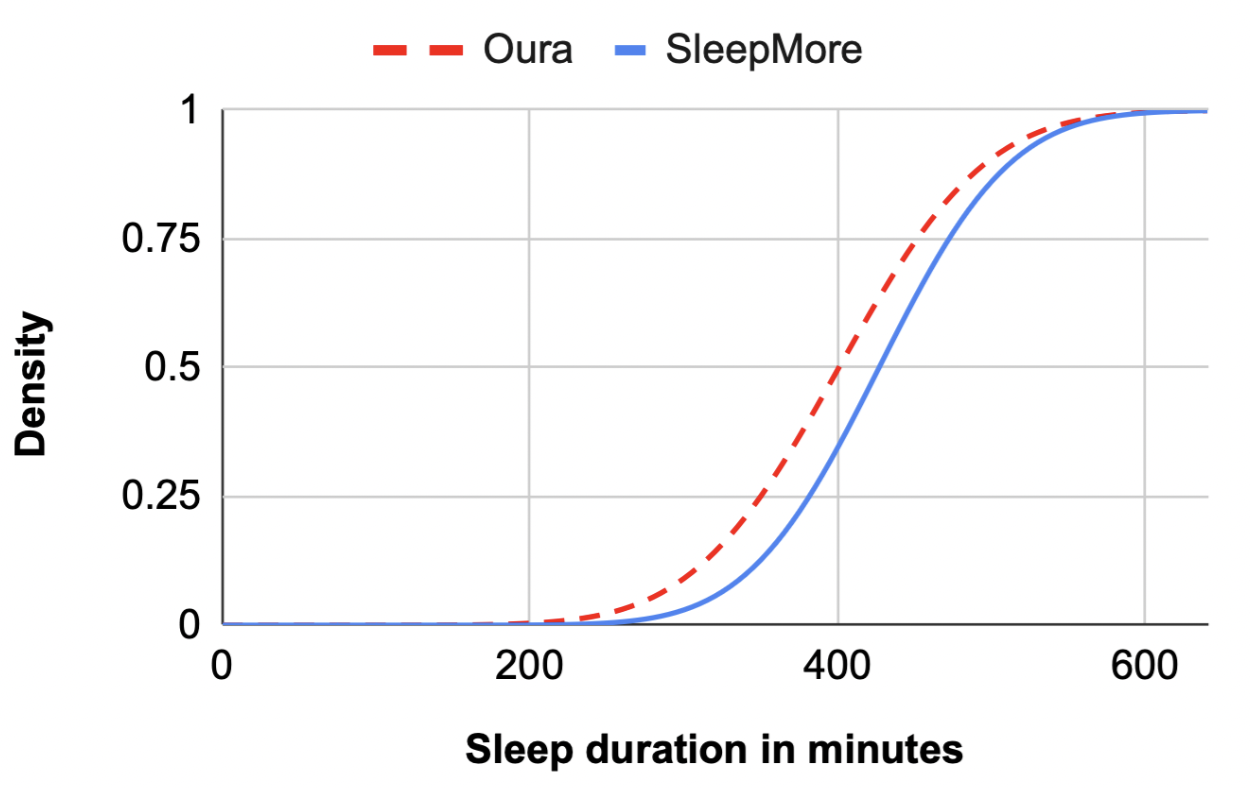}
\caption{Sleep duration CDF by Oura and \system.}
\label{fig:sleepdurationCDF}
\end{minipage}
\end{figure}

Table \ref{tab:sleepdurationCompare} summarizes the average descriptive statistics for the sleep durations measured by both the Oura ring and predicted by \system for the data collection study participants. The differences in these two distributions were statistically insignificant (with p>.1). In particular, both modalities measured that 50\% of the participants sleep for at least 400 minutes per day while the total population was measured to sleep for at least 600 minutes, as shown in Figure \ref{fig:sleepdurationCDF}.\\

\noindent \textbf{Key Takeaway.} \new{For \system to remain useful as a complementary tool in long-term sleep monitoring, our model performance must match that of the Oura ring. Our findings recommend using multiple personal devices for the system to produce the most accurate predictions.}

\subsection{Robustness Across User and Sleeping Habits}
\label{sec:robustsleep}
The following analysis seeks to understand how robust \system is to different types of users and their habitual sleep patterns. \new{As stated in Table \ref{tab:datasumm}, our student participants' average bed and wake times are 1:47 am and 8:29 am, respectively. However, there were outlying records of sleep behaviors where students reportedly went to bed at 5:30 am and woke up at 1:15 pm. }

\subsubsection{Circadian and Night Owl Phase} 
\begin{figure}[h!]
\centering
\includegraphics[width=.8\columnwidth]{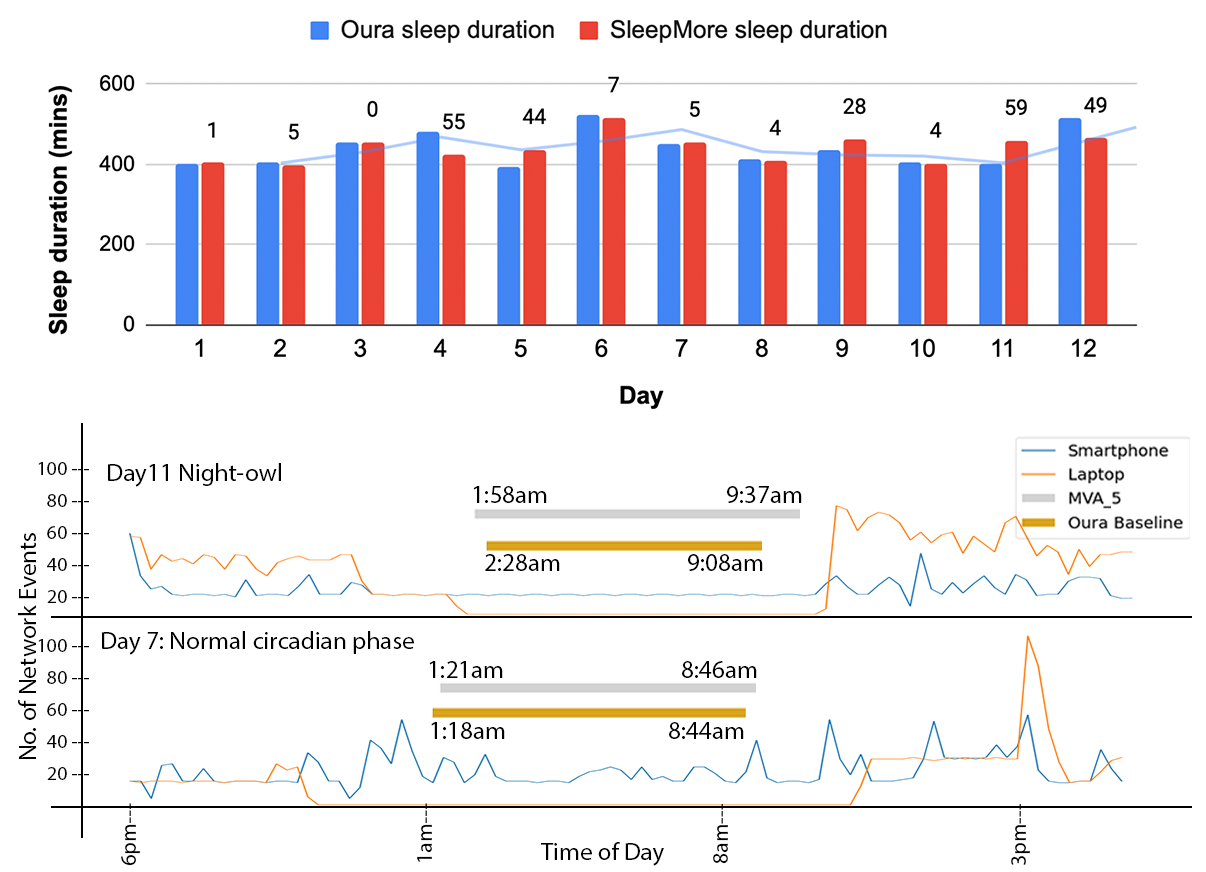}
\caption{Comparing Oura and \system for P4. Top charts sleep duration difference in minutes.}
\label{fig:sleepduration}
\end{figure}

Figure \ref{fig:sleepduration} charts participant P4, who typically slept at 1:25 am (stdev. 36 minutes) every night, corresponding to a \emph{normal} circadian phase. For the entire 12 days that P4 provided useful Oura data,  \system predicts them getting approximately 7 hours of sleep on average, thus maintaining a regular sleep pattern \cite{phillips2017irregular}, such as on Day 7. This includes Day 11, when they significantly shifted their usual sleep time – even so, \system predicted their sleep duration within 30 minutes sleep time error (and 29 minutes wake time error, totaling 59 minutes).

\begin{figure}[h!]
\centering
\includegraphics[width=1\columnwidth]{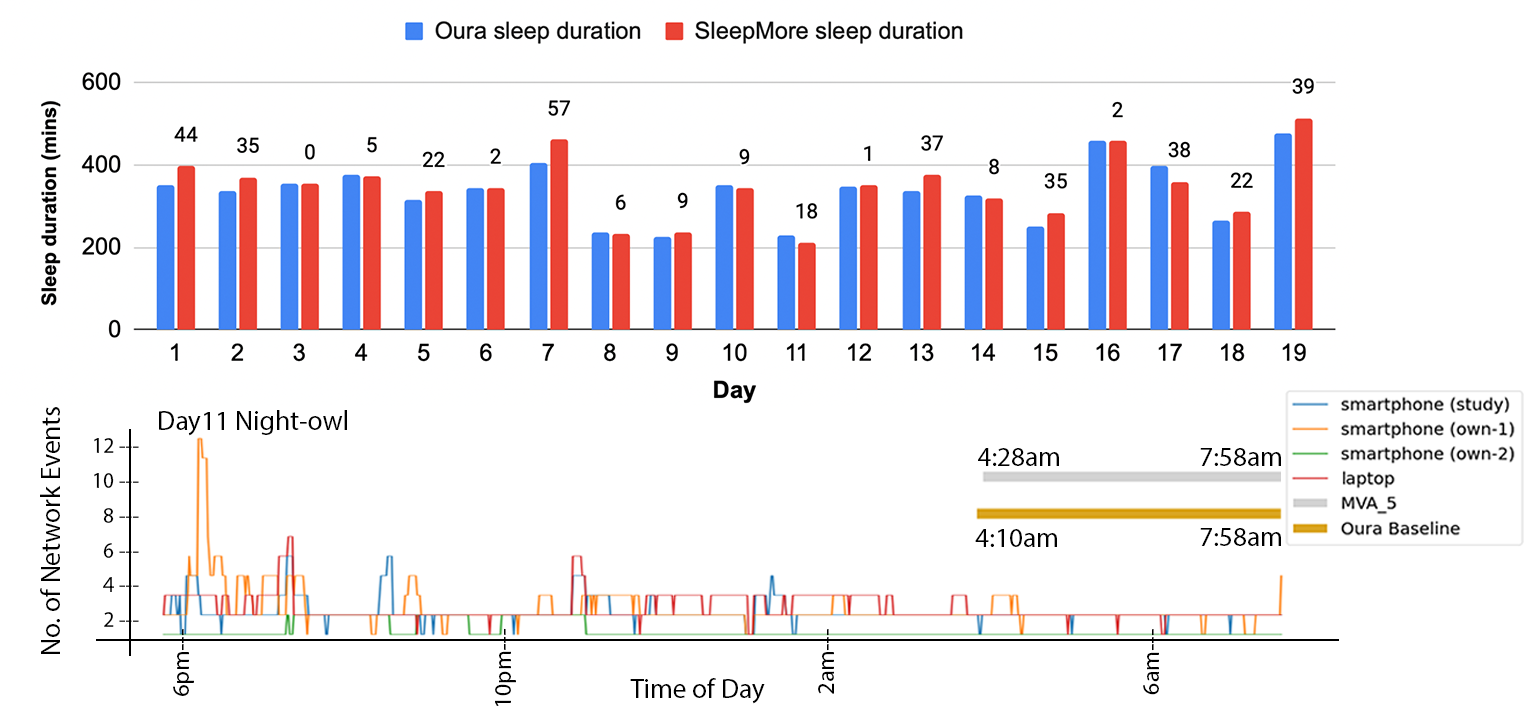}
\caption{Night owl sleep schedule by P37, identified to maintain an irregular sleep schedule.}
\label{fig:chrono2}
\end{figure}

A separate example is participant P37, as per Figure \ref{fig:chrono2}, whose average bedtime clocked at 2:09 am (stdev. 1 hour 08 minutes). On average, participant P37 received 5 hours 30 minutes (stdev. 1 hour 12 minutes) of sleep during the study period. Unlike participant P4, we identified participant P37 as an irregular sleeper \cite{taub1978behavioral}, adopting a mix of habitual \emph{night owl} \cite{mecacci1983morningness} and \emph{normal} circadian sleep schedules. On Day 11, P37 received approximately 3 hours 30 minutes of sleep. Even with this large shift, \system predicted their sleep duration with less than 18 minutes of sleep time error.

\subsubsection{Disruptive Sleep Event in an Irregular Sleeping Pattern}
\begin{figure}[h!]
\centering
\includegraphics[width=.95\columnwidth]{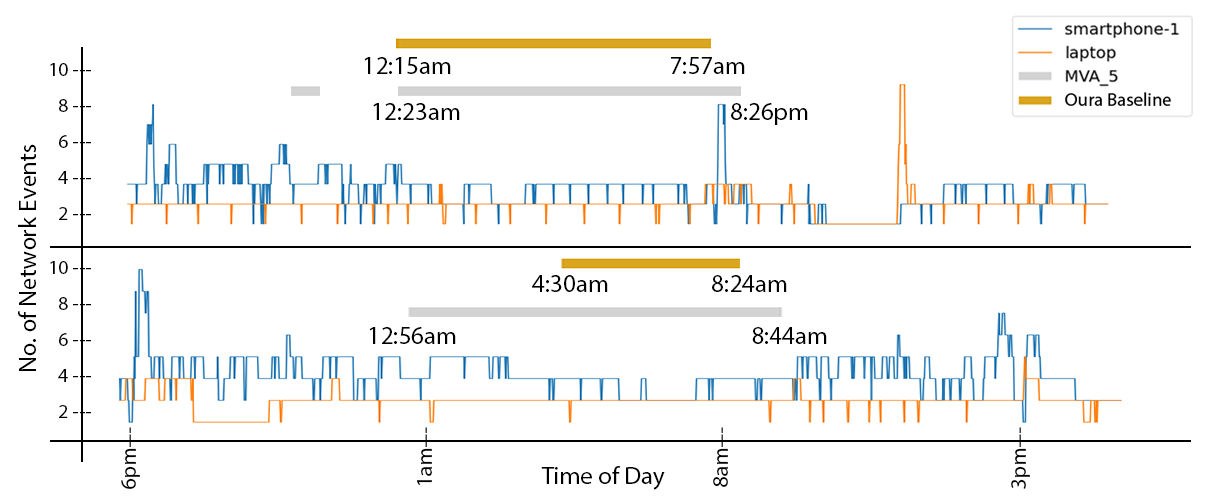}
\vspace{-0.1in}
\caption{\new{Model performance is affected by disruptive sleep event (bottom chart) for a regular sleeper, P24.}}
\label{fig:outlierP24}
\end{figure}

\begin{figure}[h!]
\centering
\includegraphics[width=.95\columnwidth]{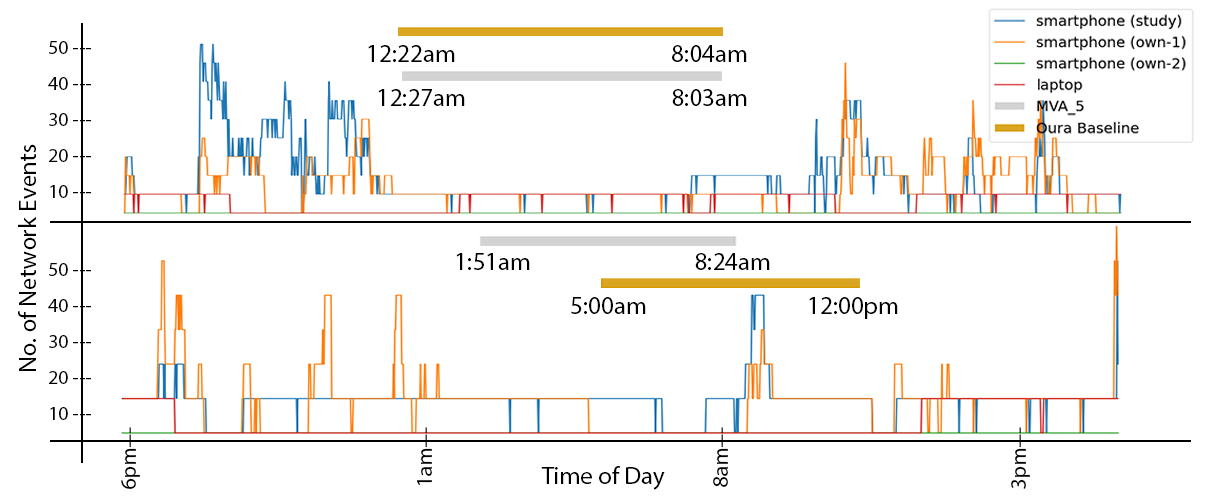}
\vspace{-0.1in}
\caption{\new{Model performance is affected by disruptive sleep event (bottom chart) for a regular sleeper, P27.}}
\label{fig:outlierP27}
\end{figure}

\new{Figures \ref{fig:outlierP24} and \ref{fig:outlierP27} show the difference in our model performance for two users, P24 and P27. Note that the disruptive sleep events (bottom chart) were each an anomaly of their habitual sleeping habits. \system can accurately predict sleep on days that users display regular sleep patterns. However, a disruptive event that leads to the user sleeping late would cause significant errors at this stage.}

\begin{figure}[h!]
\centering
\includegraphics[width=.95\columnwidth]{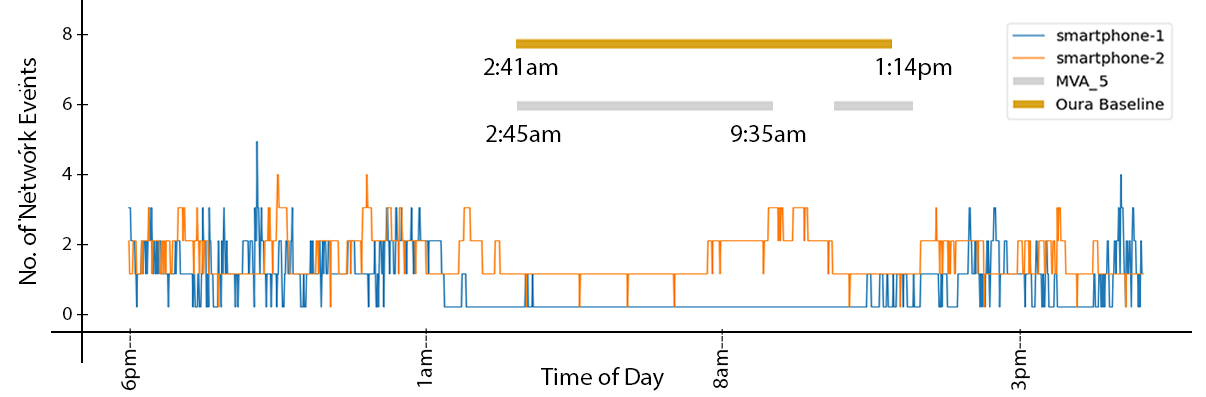}
\vspace{-0.1in}
\caption{\new{Considering the most extended sleep period perversely overlooks split-like sleep behavior for P9.}}
\label{fig:outlierWiFiAwake}
\end{figure}

A primary reason for outliers is the lack of training data, specifically for bedtime hours well past midnight and before dawn. We counted only 30 samples (equivalent to less than $\approx$ 5\%) of (all) students' data points who slept beyond 4:00 am throughout the four weeks. Additionally, these data points were mainly attributed to P37 (above), whose semi-personalized model accurately predicted their irregularities. As a result, P24 and P27's semi-personalized models had not learned to predict these outcomes. In similar cases where users did not regularly sleep in the early morning hours, the average bedtime error is within 115 minutes, and the wake time error is within 70 minutes.

\subsubsection{Wake Events Amid Nocturnal Sleep}
\new{While our system is modeled to detect nocturnal sleep, \system can handle the detection of wake states at per minute granularity from disruption of WiFi events occurring amid long sleep periods. As per Figure \ref{fig:outlierWiFiAwake} for P9, detecting continuous wake states occurring over an hour negatively affected \system's ability to estimate sleep duration that is more accurate to the ground truth. Instead, \system registered their wake time to be 9:35 am, despite a continuing sleep period between 11:00 am to 1:00 pm. Considering the longest sleep period can generally help improve overall performance (see Figures \ref{fig:smoothingProblem} and \ref{fig:outlierP24} for P2 and P24), but it can perversely overlook other sleep period that comes soon after the wake period. Note, as we explained in Section \ref{sec:study}, our sleep baseline only considers nocturnal sleep (meaning a split-like sleep behavior would be consider these split events as a whole).}\\

\noindent \textbf{Key Takeaway.} Indeed, a significant challenge foreseen in our approach is determining a user's bedtime and wake time by monitoring coarse-grained WiFi features. Our estimation method values sleep states made in the last 5 minutes to predict each user's upcoming sleep/awake state and takes the most extended sleep period as the nocturnal sleep duration. \new{The system is robust to regular and irregular sleepers. However, it does not handle unexpected disruptive sleep events from regular sleepers as well, primarily since it currently lacks samples of users sleeping at dawn. Further, considering the most extended sleep period can impact our model performance, especially among users who exhibit split-like sleep behavior.}

\subsection{Home-Use Applicability}
\label{sec:home}
These evaluations thus far have focused on our primary user study among on-campus student residents. One question that remains unaddressed is \system's performance in a private home setting. \new{In what follows, we tested \system in a cross-environment. That is, the semi-personalized model for each home participant is built using students' WiFi sleep dataset (the primary study) but with 40\% of their data.}

\begin{figure}[h!]
\begin{minipage}{0.42\textwidth}
\captionsetup{justification=centering}
\centering
\vspace{.8\baselineskip}
\includegraphics[width=\columnwidth]{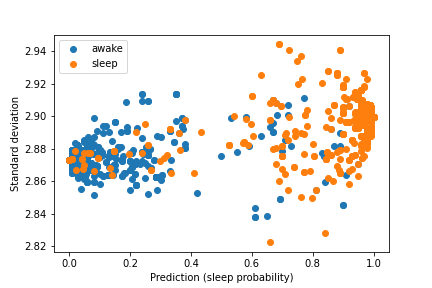}
\vspace{.1\baselineskip}
\caption{Classification of sleep states and measure of dispersion for $HomeUser_{1}$.}
\label{fig:homesleepClassification}
\end{minipage}
\hspace{0.01cm}
\begin{minipage}{0.56\textwidth}
\vspace{2\baselineskip}
\captionsetup{justification=centering}
\centering
\captionof{table}{\new{\system's performance and errors all within 1\% uncertainty rates in a private home setting.}}
\scalebox{.9}{
\begin{tabular}{l|c|c|c}\hline
& \textbf{$HomeUser_{1}$} & \textbf{$HomeUser_{2}$} & \textbf{$HomeUser_{3}$} \\ \hline
\textbf{Accuracy} & 0.986 & 0.988 & 0.991 \\
\textbf{Precision} & 0.977 & 0.983 & 0.986 \\
\textbf{Recall} & 0.975 & 0.978 & 0.982 \\
\textbf{F1} & 0.976 & 0.980 & 0.984 \\ \hline
\textbf{$T_{sleep}$ (<=1\%)} & \makecell[c]{6-8 mins\\(mean: 6 mins)}  & \makecell[c]{4-11 mins\\(mean: 6 mins)} & \makecell[c]{1-17 mins\\(mean: 6 mins) } \\ \hline
\textbf{$T_{wake}$ (<=1\%)} & \makecell[c]{1-28 mins\\(mean: 8 mins)} & \makecell[c]{1-27 mins\\(mean: 7 mins)} & \makecell[c]{1- 4 mins\\(mean: 2 mins) } \\ \hline
\textbf{$T_{sleep}$ (<=5\%)} & \makecell[c]{2-5 mins\\(mean: 3 mins)}  & \makecell[c]{24 mins\\(1 sample)} & - \\ \hline
\textbf{$T_{wake}$ (<=5\%)} & \makecell[c]{3-19 mins\\(mean: 8 mins)} & \makecell[c]{6 mins\\(1 sample)} & - \\ \hline
\end{tabular} }  
\label{tab:modelHome}
\end{minipage}
\end{figure}

Figure \ref{fig:homesleepClassification} plots the classification of sleep and awake states using their semi-personalized random forest model for $HomeUser_{1}$. Respectively, we plot the standard deviation of each prediction (square root of the variance), giving us a better idea of the disparity of data from one another. As summarized in Table \ref{tab:modelHome}, our model achieves an overall performance of 98.6\% accuracy and a high 97.5\% recall in predicting sleep states. The effect of these predictions leads to our estimation model producing \new{between 2 to 8  minutes of sleep error} and \new{1 to 28} minutes of wake error. \new{the evaluation yielded similar results for our home users sharing the same residence.} Note that all predictions were estimated within a 3\% uncertainty rate; therefore, no outliers were removed.

\begin{figure}[h!]
\begin{minipage}{0.55\textwidth}
\captionsetup{justification=centering}
\vspace{-0.15in}
\centering
\includegraphics[width=\textwidth,height=4cm]{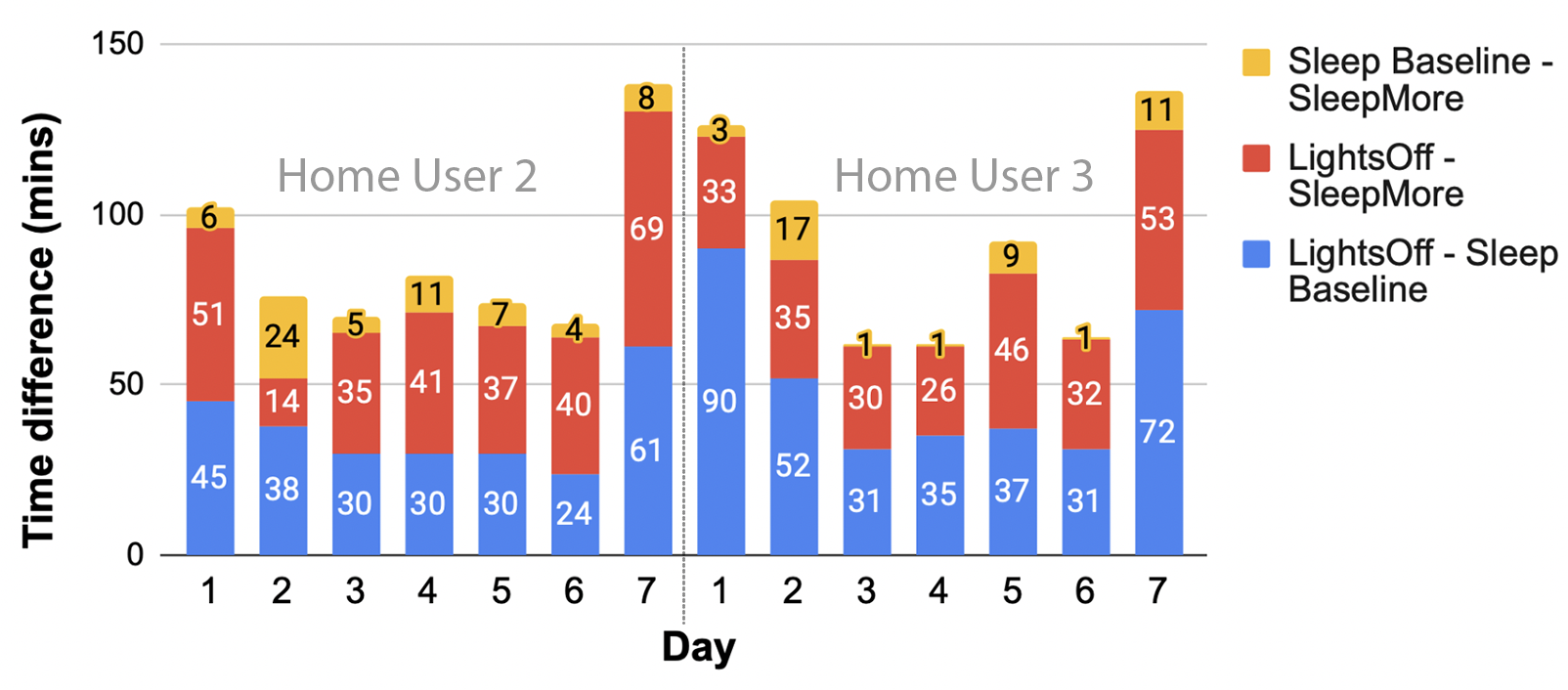}
\vspace{-0.3in}
\caption{\new{Time difference of \system with lights off and sleep baseline.}}
\label{fig:homelightcomparison}
\end{minipage}
\hspace{0.01cm}
\begin{minipage}{0.4\textwidth}
\captionsetup{justification=centering}
\centering
\captionof{table}{\new{Performance difference with and without shared device features.}}
\scalebox{.9}{
\begin{tabular}{l|c|c}\hline
& \textbf{$HomeUser_{2}$} & \textbf{$HomeUser_{3}$} \\ \hline
\textbf{Accuracy} & 0.991 (+1.9\%) & 0.993 (+0.2\%) \\
\textbf{Precision} & 0.991 (+0.8\%) & 0.989 (+0.3\%)\\
\textbf{Recall} & 0.978 (+0\%) & 0.986 (+0.4\%)\\
\textbf{F1} & 0.984 (+0.4\%) & 0.988 (+0.4\%) \\ \hline
\textbf{$T_{sleep}$ (<=1\%)} & \makecell[c]{2-7 mins\\(mean: 4 mins)}  & \makecell[c]{5-8 mins\\(mean: 6 mins)} \\ \hline
\textbf{$T_{wake}$ (<=1\%)} & \makecell[c]{1-26 mins\\(mean: 6 mins)}  & \makecell[c]{1-8 mins\\(mean: 4 mins)} \\ \hline
\end{tabular} }  
\label{tab:sharedDevice}
\end{minipage}
\end{figure}

\new{In a traditional sleep setting, `lights off' signals a person's intention to sleep. However, the intention to sleep to actual sleep time (sleep latency) is one measure that varies across persons. To examine potential discrepancies, we compared the time difference produced by our system with $HomeUser_{2}$ and $HomeUser_{3}$'s lights off logs and sleep baseline. As per Figure \ref{fig:homelightcomparison}, using `lights off' as a point of reference for sleep produced between 24-90 minutes time difference for our participants compared to the Fitbit wearable, with the largest differences occurring over the weekend. The time difference between our system and `lights off' ranges from 14-69 minutes. It should be noted that while our system cannot detect sleep onset latency with coarse-grained information, the time difference is considerably less against the Fitbit compared to lights off.}

\subsubsection{Utility of Smart Home Devices}
\new{These results thus far have shown promise in utilizing WiFi data from personal devices. We emphasize personal devices because much work has proven that these device types are a reasonable proxy for estimating user behavior. However, with the increasing use of smart home devices today, we briefly investigate their impact on our model performance. Smart home devices, although shared, are almost guaranteed to be connected to home WiFi networks. Table \ref{tab:sharedDevice} presents the performance difference for $HomeUser_{2}$ and $HomeUser_{3}$, respectively. In removing features generated by smart home devices, our model performance improves overall accuracy but with insignificant difference (p>.1). Note, however, no one prediction was made beyond 1\% uncertainty rate.}

\begin{figure}[h!]
\centering
\includegraphics[width=.9\columnwidth]{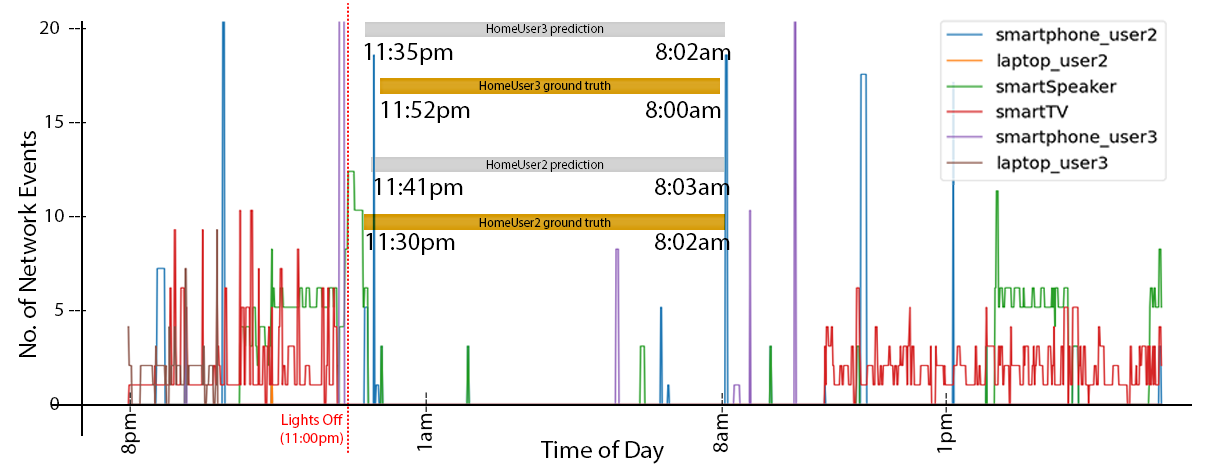}
\vspace{-0.2in}
\caption{\new{Model performance on device usage for $HomeUser_{2}$ and $HomeUser_{3}$.}}
\label{fig:homecombined}
\end{figure}

\new{It is essential to highlight the following attributes in considering the relevance of our results. First, for many people, the second-nature action of switching off the TV every night before bed makes it more reasonable to approximate user states. Equally, this behavior may not be true for users who habitually leave their TV on during bedtime. In this case, our users are a couple with similar sleep hygiene who did not exhibit a pattern of leaving their devices running through the night. Second, they displayed similar sleep schedules, with one person shortly sleeping and waking up after another. Monitoring WiFi network data from their smart TV and speaker made little impact on our model performance, whether at night or in the morning. Where this condition does not hold, monitoring WiFi data from shared devices will be less straightforward for sleep monitoring.}

\new{One way to accommodate the nuances in users' digital device practices is by distinguishing each device as primary, secondary, or shared use, assigning weights based on utilization frequency and time of day. By operationalizing feature extraction in such a way, we could more accurately determine the user utilizing a shared device. For example, Figure \ref{fig:homecombined} charts the combined use of devices for $HomeUser_{2}$ and $HomeUser_{3}$. While both users shared the same device practice through the night, it is unclear who was operating the smartTV, and smart speaker from 10 am onwards. For \system to extend its capability in estimating daytime naps, detecting who is awake while utilizing the device during the day will be crucial to our model development.}\\

\noindent \textbf{Key Takeaway} The primary objective of this secondary study is to demonstrate the applicability of our system in a private home setting, albeit on a different time scale and different sleep baseline. While the insights we derived were based on a limited observation window, we show how our technique is adaptable to different residential settings as long as users connect their devices to a dedicated WiFi network when at home. In a shared setting, we distinguish a user based on their personal devices' MAC addresses and can singularize different occupant's behavior based on their individual list of personal devices. 

\subsection{Performance Comparison against Prior Work}
Our final analysis compares \system's performance against a system implementation of three unsupervised learning techniques \new{by prior work \cite{mammen2021wisleep}. Similarly, we evaluate the performance of these models using leave-one-out cross validation}.

\begin{table}[h!]
\centering
\caption{Comparison with prior techniques and our sleep estimation errors within 5\% uncertainty rate.}
\scalebox{.9}{
\begin{tabular}{l|c|c|c|c|c|c}\hline
\textbf{Technique} & \textbf{Acc} & \textbf{Prec} & \textbf{Rec} & \textbf{F1} & \textbf{$T_{sleep}$ error in mins} & \textbf{$T_{wake}$ error in mins} \\ \hline
\system & 0.939 & 0.896 & 0.905 & 0.900 & 15-28 (<= 5\%) & 7-29 (<= 5\%) \\
Ensemble based & 0.787 & 0.586 & 0.891 & 0.707 & 139 & 175 \\
Norm. Prior & 0.799 & 0.600 & 0.919 & 0.726  & 190 & 109 \\
Hier. Prior & 0.814& 0.622 & 0.914 & 0.740  & 193 & 111 \\
\hline
\end{tabular} }     
\label{tab:modelComparison}
\end{table}

The comparison drawn in prior work was based on three Bayesian change point detection methods using WiFi device activity from all devices (i.e., \emph{all\_networkEvents} feature), looking at the rate of change of values in a single variable time-series data; normal prior \cite{el2018sleep}, hierarchical prior \cite{cuttone2017sensiblesleep}, and an ensemble of normal, uniform, and hierarchical priors \cite{mammen2021wisleep}. The authors found that techniques using only a hierarchical prior or a normal prior are useful when users have regular sleep patterns and the data is not subjected to too much noise. However, basic profiling of users is required to best decide on these priors. An ensemble model accommodates the irregular sleepers, including a uniform prior model, the normal prior model, and the hierarchical prior model.

\new{Building on these findings, we employed the system developed by Mammen \etal \cite{mammen2021wisleep} on our primary study dataset.} Table \ref{tab:modelComparison} summarizes the performance comparison based on our primary study (student participants' data) and shows that \system achieves significantly better accuracy, recall, and precision (p<.01). A plausible explanation for the change point detection models failing is the absence of location information. Specifically, the amount of WiFi network activity we calculate every minute does not consider changes between places and only checks for a coarse residential location. The above change point detection techniques are suitable in situations where we have no access to the training data and want a high-level understanding of users' behavior in a residential location. However, when we want to do more fine-grained analysis and have access to training data from multiple devices, \system can provide more accuracy. \\

\noindent \textbf{Key Takeaway.} While earlier work similarly utilizes WiFi connection data to predict sleep, two main differences set our work apart: First, these works relied solely on single-device monitoring through users' smartphones. Second, change point detection is employed in prior work to determine sleep from changes in WiFi event rates in different locations.
\section{Discussion and Limitations}
\label{sec:discussion}
Our study's objectives were to develop a sleep prediction solution that can accurately predict sleep by monitoring WiFi device activity for multi-user devices. Here we discuss the implications of our findings.

\subsection{Assessing Type of Sleep Characteristics}
We provide evidence of our technique producing key sleep characteristics comparable to the Oura ring, such as bedtime, wake time, and sleep duration. A clear limitation of utilizing a coarse-grained data source is the inability to differentiate the four sleep stages (i.e., Deep, Light, REM, wake) \new{and sleep onset latency}. \new{Currently, only estimation of bedtime and wake time is possible with \system.} Hence, it should be emphasized that our technique aims not to replace existing sleep-sensing modalities that offer fine-grained information but instead to complement the use of such modalities in supporting sleep monitoring, especially over longitudinal periods. At present, the training for our model is notably restricted to nocturnal sleep cycles. \new{The operational challenges faced in our user study (see Section \ref{sec:study}) hindered our progress in experimenting with predicting nap times (note: today, the software update in wearable sleep trackers are equipped with nap detection). With \system's capability to detect awake and sleep states at per minute granularity, the system could conceivably detect different sleep patterns, including polyphasic and split-like sleep behavior. Our estimation technique, at present, smooths out wake events amid a sleep period at a fixed threshold, thus requires implementing dynamic thresholding in setting naps and split-like sleep behavior for different users.}

\subsection{Extending to Different User Populations and Device Types}
\new{The main population target for this study is students and a smaller scale of full-time working professionals with regular work schedules. However, different populations might exhibit different sleep patterns. While students have the flexibility to adjust their sleep schedule to academic/social demands by sleeping during the day, most daytime workers have to follow nighttime sleep. Sleep patterns might be highly irregular in shift workers or patients with insomnia, which our model development has not accounted for in this phase. }

\new{Nonetheless, the cross-environment model development of on-campus student residents to home participants exemplifies the feasibility of supporting large-scale sleep monitoring in family dwelling units. Although the supplementary home study was a minor scale, it should be noted that our home participants were not in a single-living environment.} Thus, it is feasible to extend sleep prediction for other family members/occupants by singularizing each user's sleep predictions based on their dedicated list of personal devices. \new{We emphasize `personal devices' because these devices (in our study and similar prior work) have shown to effectively approximate user presence and activity.}

\new{We learned from our home study the promise of monitoring WiFi data generated from shared information appliances such as smart TV and smart speakers, which are almost guaranteed to be connected to home WiFi networks. However, the practicality of using these devices grows with increasing complexity when more people share the device. It will also pose as an issue when extending \system to detect daytime naps, requiring the model to learn the user more likely to be using the device based on their routines. This is where the personalization of device feature weights may be relevant.}

\subsection{Improvements to Handle Corner Cases} 
\begin{figure}[h!]
\centering
\includegraphics[width=.7\columnwidth]{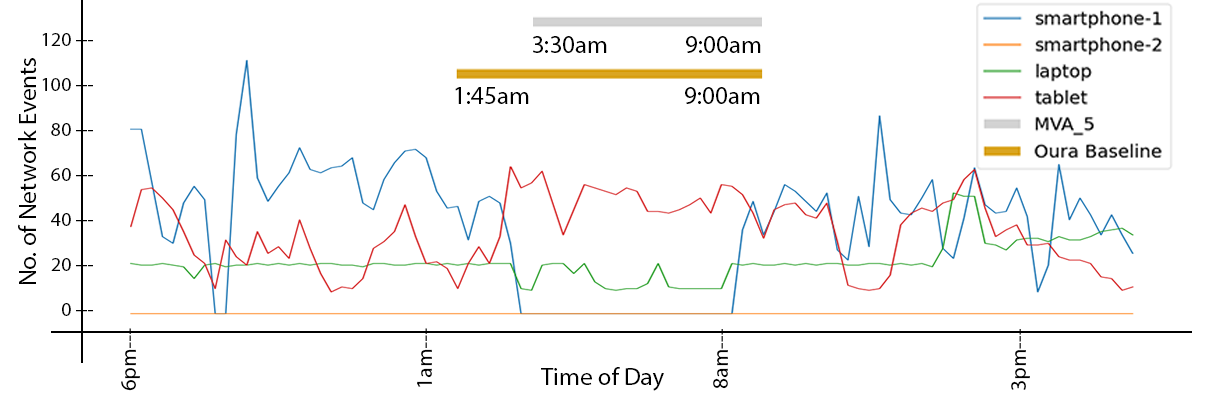}
\vspace{-1\baselineskip}
\caption{Outlier with large error.}
\label{fig:cornercase}
\end{figure}

Finally, Figure \ref{fig:cornercase} exemplifies an error our system encounters that could be attributed to multiple variables and conditions. \new{In fact, it is common for devices to stay connected to the WiFi for activities such as software updates, music/video streaming during the night. As previously suggested, one possibility to reduce the impact of these device features in sleep prediction is assigning different weights to each device as a separate feature.}

\new{Further, under conditions where \system is utilized as a complementary tool over long-term observations, a feasible way to correct sleep predictions with large errors is by supplementing historical data from another sleep tracker modality to train a user's semi-personalized model continuously. This is also where our work continues with the LSTM implementation, briefly presented in Section \ref{sec:resultDL}. Months of training data will allow us to experiment with building a classifier that compares sleep prediction the following day to longer historical sequences. Instead of using WiFi network event features extracted per day, we can couple our model input with WiFi network data of users over weekdays, weekends, and calendar events, tailored to their device habits and sleep routines. Current work relied on the trial-and-error method for model architecture and hyperparameters. Extension to this effort can include employing more sophisticated optimization approaches for better model configuration.}

 \section{Conclusions and Future Work}
This work proposed \system as a \new{promising, low cost and easily scalable supplementary solution to commercial sleep trackers for long-term} sleep monitoring in residences such as dormitories and homes. \system uses a supervised learning and estimation model that leverages WiFi device activity, collected directly from the WiFi infrastructure from multiple user devices to predict sleep. It does this in two steps; First, it determines if a user is in a sleep or awake state every minute using a random forest semi-personalized model. For each state, the model employs an infinitesimal jackknife variance estimation method to calculate the confidence for each prediction, noting down every state with low confidence. Second, it processes these sequences of sleep and awake, using a moving average to estimate the user's bedtime and wake times. It determines the uncertainty rate of a nocturnal sleep estimate instanced by the number of low confidence states present in this sequence. Our validation used data collected from 46 participants living in on-campus dormitories, predicting sleep states with a high recall of $\approx$ 90\% recall. This performance renders sleep duration estimations within 15-28 minutes sleep time error and 7-29 minutes wake time error, henceforth proving statistically significant improvements over prior work. \new{We also demonstrated the application of \system in private home settings and provide insights into the utilization of shared home devices.} A detailed comparative analysis highlighted the importance of multiple user devices to accurately estimate sleep duration and the robustness of \system at predicting sleep measures across different users with different sleep patterns and regularities. As future work, we plan to incorporate \system into a scalable solution for on-campus health and behavioral risk surveillance to help college students at risk of sleep deprivation and related issues.
\begin{acks}
We thank Alyssa Ng and Xin Yu Chua of National University of Singapore for assisting in participant recruitment and communication during the study and Bryce Parkman of University of Massachusetts Amherst for helping with data preprocessing. Thank you to Dr. Ju Lynn Ong of National University of Singapore for providing insight and expertise in guiding our research. This research is supported by National Science Foundation grants 1763834, 1836752, 2021693, 2020888, 2105494, Support Funds for the Centre for Sleep and Cognition, National University of Singapore, and the Singapore Management University Lee Kong Chian Fellowship.
\end{acks}

\bibliographystyle{ACM-Reference-Format}
\bibliography{citations}


\begin{thebibliography}{61}


\ifx \showCODEN    \undefined \def \showCODEN     #1{\unskip}     \fi
\ifx \showDOI      \undefined \def \showDOI       #1{#1}\fi
\ifx \showISBNx    \undefined \def \showISBNx     #1{\unskip}     \fi
\ifx \showISBNxiii \undefined \def \showISBNxiii  #1{\unskip}     \fi
\ifx \showISSN     \undefined \def \showISSN      #1{\unskip}     \fi
\ifx \showLCCN     \undefined \def \showLCCN      #1{\unskip}     \fi
\ifx \shownote     \undefined \def \shownote      #1{#1}          \fi
\ifx \showarticletitle \undefined \def \showarticletitle #1{#1}   \fi
\ifx \showURL      \undefined \def \showURL       {\relax}        \fi
\providecommand\bibfield[2]{#2}
\providecommand\bibinfo[2]{#2}
\providecommand\natexlab[1]{#1}
\providecommand\showeprint[2][]{arXiv:#2}

\bibitem[\protect\citeauthoryear{Abadi, Agarwal, Barham, Brevdo, Chen, Citro,
  Corrado, Davis, Dean, Devin, et~al\mbox{.}}{Abadi et~al\mbox{.}}{2016}]%
        {abadi2016tensorflow}
\bibfield{author}{\bibinfo{person}{Mart{\'\i}n Abadi}, \bibinfo{person}{Ashish
  Agarwal}, \bibinfo{person}{Paul Barham}, \bibinfo{person}{Eugene Brevdo},
  \bibinfo{person}{Zhifeng Chen}, \bibinfo{person}{Craig Citro},
  \bibinfo{person}{Greg~S Corrado}, \bibinfo{person}{Andy Davis},
  \bibinfo{person}{Jeffrey Dean}, \bibinfo{person}{Matthieu Devin},
  {et~al\mbox{.}}} \bibinfo{year}{2016}\natexlab{}.
\newblock \showarticletitle{Tensorflow: Large-scale machine learning on
  heterogeneous distributed systems}.
\newblock \bibinfo{journal}{\emph{arXiv preprint arXiv:1603.04467}}
  (\bibinfo{year}{2016}).
\newblock


\bibitem[\protect\citeauthoryear{Abdullah, Matthews, Murnane, Gay, and
  Choudhury}{Abdullah et~al\mbox{.}}{2014}]%
        {abdullah2014towards}
\bibfield{author}{\bibinfo{person}{Saeed Abdullah}, \bibinfo{person}{Mark
  Matthews}, \bibinfo{person}{Elizabeth~L Murnane}, \bibinfo{person}{Geri Gay},
  {and} \bibinfo{person}{Tanzeem Choudhury}.} \bibinfo{year}{2014}\natexlab{}.
\newblock \showarticletitle{Towards circadian computing: ``early to bed and
  early to rise'' makes some of us unhealthy and sleep deprived}. In
  \bibinfo{booktitle}{\emph{Proceedings of the 2014 ACM international joint
  conference on pervasive and ubiquitous computing}}.
  \bibinfo{pages}{673--684}.
\newblock


\bibitem[\protect\citeauthoryear{Altevogt, Colten, et~al\mbox{.}}{Altevogt
  et~al\mbox{.}}{2006}]%
        {altevogt2006sleep}
\bibfield{author}{\bibinfo{person}{Bruce~M Altevogt}, \bibinfo{person}{Harvey~R
  Colten}, {et~al\mbox{.}}} \bibinfo{year}{2006}\natexlab{}.
\newblock \showarticletitle{Sleep disorders and sleep deprivation: an unmet
  public health problem}.
\newblock  (\bibinfo{year}{2006}).
\newblock


\bibitem[\protect\citeauthoryear{Altini and Kinnunen}{Altini and
  Kinnunen}{2021}]%
        {altini2021promise}
\bibfield{author}{\bibinfo{person}{Marco Altini} {and} \bibinfo{person}{Hannu
  Kinnunen}.} \bibinfo{year}{2021}\natexlab{}.
\newblock \showarticletitle{The promise of sleep: A multi-sensor approach for
  accurate sleep stage detection using the oura ring}.
\newblock \bibinfo{journal}{\emph{Sensors}} \bibinfo{volume}{21},
  \bibinfo{number}{13} (\bibinfo{year}{2021}), \bibinfo{pages}{4302}.
\newblock


\bibitem[\protect\citeauthoryear{Bartel, Scheeren, and Gradisar}{Bartel
  et~al\mbox{.}}{2019}]%
        {bartel2019altering}
\bibfield{author}{\bibinfo{person}{K Bartel}, \bibinfo{person}{R Scheeren},
  {and} \bibinfo{person}{M Gradisar}.} \bibinfo{year}{2019}\natexlab{}.
\newblock \showarticletitle{Altering adolescents' pre-bedtime phone use to
  achieve better sleep health}.
\newblock \bibinfo{journal}{\emph{Health communication}} \bibinfo{volume}{34},
  \bibinfo{number}{4} (\bibinfo{year}{2019}), \bibinfo{pages}{456--462}.
\newblock


\bibitem[\protect\citeauthoryear{Bovornkeeratiroj, Wamburu, Irwin, and
  Shenoy}{Bovornkeeratiroj et~al\mbox{.}}{2021}]%
        {bovornkeeratiroj2021vpeak}
\bibfield{author}{\bibinfo{person}{Phuthipong Bovornkeeratiroj},
  \bibinfo{person}{John Wamburu}, \bibinfo{person}{David Irwin}, {and}
  \bibinfo{person}{Prashant Shenoy}.} \bibinfo{year}{2021}\natexlab{}.
\newblock \showarticletitle{VPeak: exploiting volunteer energy resources for
  flexible peak shaving}. In \bibinfo{booktitle}{\emph{Proceedings of the 8th
  ACM International Conference on Systems for Energy-Efficient Buildings,
  Cities, and Transportation}}. \bibinfo{pages}{121--130}.
\newblock


\bibitem[\protect\citeauthoryear{Cappuccio, Cooper, D'Elia, Strazzullo, and
  Miller}{Cappuccio et~al\mbox{.}}{2011}]%
        {cappuccio2011sleep}
\bibfield{author}{\bibinfo{person}{Francesco~P Cappuccio},
  \bibinfo{person}{Daniel Cooper}, \bibinfo{person}{Lanfranco D'Elia},
  \bibinfo{person}{Pasquale Strazzullo}, {and} \bibinfo{person}{Michelle~A
  Miller}.} \bibinfo{year}{2011}\natexlab{}.
\newblock \showarticletitle{Sleep duration predicts cardiovascular outcomes: a
  systematic review and meta-analysis of prospective studies}.
\newblock \bibinfo{journal}{\emph{European heart journal}}
  \bibinfo{volume}{32}, \bibinfo{number}{12} (\bibinfo{year}{2011}),
  \bibinfo{pages}{1484--1492}.
\newblock


\bibitem[\protect\citeauthoryear{Chen, Xiao, Tang, Fernandez-Mendoza, and
  Cao}{Chen et~al\mbox{.}}{2021}]%
        {chen2021apneadetector}
\bibfield{author}{\bibinfo{person}{Xianda Chen}, \bibinfo{person}{Yifei Xiao},
  \bibinfo{person}{Yeming Tang}, \bibinfo{person}{Julio Fernandez-Mendoza},
  {and} \bibinfo{person}{Guohong Cao}.} \bibinfo{year}{2021}\natexlab{}.
\newblock \showarticletitle{ApneaDetector: Detecting Sleep Apnea with
  Smartwatches}.
\newblock \bibinfo{journal}{\emph{Proceedings of the ACM on Interactive,
  Mobile, Wearable and Ubiquitous Technologies}} \bibinfo{volume}{5},
  \bibinfo{number}{2} (\bibinfo{year}{2021}), \bibinfo{pages}{1--22}.
\newblock


\bibitem[\protect\citeauthoryear{Chen, Lin, Chen, Lane, Cardone, Wang, Li,
  Chen, Choudhury, and Campbell}{Chen et~al\mbox{.}}{2013}]%
        {chen2013unobtrusive}
\bibfield{author}{\bibinfo{person}{Zhenyu Chen}, \bibinfo{person}{Mu Lin},
  \bibinfo{person}{Fanglin Chen}, \bibinfo{person}{Nicholas~D Lane},
  \bibinfo{person}{Giuseppe Cardone}, \bibinfo{person}{Rui Wang},
  \bibinfo{person}{Tianxing Li}, \bibinfo{person}{Yiqiang Chen},
  \bibinfo{person}{Tanzeem Choudhury}, {and} \bibinfo{person}{Andrew~T
  Campbell}.} \bibinfo{year}{2013}\natexlab{}.
\newblock \showarticletitle{Unobtrusive sleep monitoring using smartphones}. In
  \bibinfo{booktitle}{\emph{2013 7th International Conference on Pervasive
  Computing Technologies for Healthcare and Workshops}}. IEEE,
  \bibinfo{pages}{145--152}.
\newblock


\bibitem[\protect\citeauthoryear{Crompton and Burke}{Crompton and
  Burke}{2018}]%
        {crompton2018use}
\bibfield{author}{\bibinfo{person}{Helen Crompton} {and} \bibinfo{person}{Diane
  Burke}.} \bibinfo{year}{2018}\natexlab{}.
\newblock \showarticletitle{The use of mobile learning in higher education: A
  systematic review}.
\newblock \bibinfo{journal}{\emph{Computers \& Education}}
  \bibinfo{volume}{123} (\bibinfo{year}{2018}), \bibinfo{pages}{53--64}.
\newblock


\bibitem[\protect\citeauthoryear{Cuttone, B{\ae}kgaard, Sekara, Jonsson,
  Larsen, and Lehmann}{Cuttone et~al\mbox{.}}{2017}]%
        {cuttone2017sensiblesleep}
\bibfield{author}{\bibinfo{person}{Andrea Cuttone}, \bibinfo{person}{Per
  B{\ae}kgaard}, \bibinfo{person}{Vedran Sekara}, \bibinfo{person}{H{\aa}kan
  Jonsson}, \bibinfo{person}{Jakob~Eg Larsen}, {and} \bibinfo{person}{Sune
  Lehmann}.} \bibinfo{year}{2017}\natexlab{}.
\newblock \showarticletitle{Sensiblesleep: A bayesian model for learning sleep
  patterns from smartphone events}.
\newblock \bibinfo{journal}{\emph{PloS one}} \bibinfo{volume}{12},
  \bibinfo{number}{1} (\bibinfo{year}{2017}), \bibinfo{pages}{e0169901}.
\newblock


\bibitem[\protect\citeauthoryear{De~Zambotti, Cellini, Goldstone, Colrain, and
  Baker}{De~Zambotti et~al\mbox{.}}{2019}]%
        {de2019wearable}
\bibfield{author}{\bibinfo{person}{Massimiliano De~Zambotti},
  \bibinfo{person}{Nicola Cellini}, \bibinfo{person}{Aimee Goldstone},
  \bibinfo{person}{Ian~M Colrain}, {and} \bibinfo{person}{Fiona~C Baker}.}
  \bibinfo{year}{2019}\natexlab{}.
\newblock \showarticletitle{Wearable sleep technology in clinical and research
  settings}.
\newblock \bibinfo{journal}{\emph{Medicine and science in sports and exercise}}
  \bibinfo{volume}{51}, \bibinfo{number}{7} (\bibinfo{year}{2019}),
  \bibinfo{pages}{1538}.
\newblock


\bibitem[\protect\citeauthoryear{de~Zambotti, Rosas, Colrain, and
  Baker}{de~Zambotti et~al\mbox{.}}{2019}]%
        {de2019sleep}
\bibfield{author}{\bibinfo{person}{Massimiliano de Zambotti},
  \bibinfo{person}{Leonardo Rosas}, \bibinfo{person}{Ian~M Colrain}, {and}
  \bibinfo{person}{Fiona~C Baker}.} \bibinfo{year}{2019}\natexlab{}.
\newblock \showarticletitle{The sleep of the ring: comparison of the OURA sleep
  tracker against polysomnography}.
\newblock \bibinfo{journal}{\emph{Behavioral sleep medicine}}
  \bibinfo{volume}{17}, \bibinfo{number}{2} (\bibinfo{year}{2019}),
  \bibinfo{pages}{124--136}.
\newblock


\bibitem[\protect\citeauthoryear{El-Khadiri, Corona, Rose, and
  Charpillet}{El-Khadiri et~al\mbox{.}}{2018}]%
        {el2018sleep}
\bibfield{author}{\bibinfo{person}{Yassine El-Khadiri},
  \bibinfo{person}{Gabriel Corona}, \bibinfo{person}{C{\'e}dric Rose}, {and}
  \bibinfo{person}{Fran{\c{c}}ois Charpillet}.}
  \bibinfo{year}{2018}\natexlab{}.
\newblock \showarticletitle{Sleep Activity Recognition Using Binary Motion
  Sensors}. In \bibinfo{booktitle}{\emph{2018 IEEE 30th International
  Conference on Tools with Artificial Intelligence (ICTAI)}}. IEEE,
  \bibinfo{pages}{265--269}.
\newblock


\bibitem[\protect\citeauthoryear{Exelmans and Van~den Bulck}{Exelmans and
  Van~den Bulck}{2016}]%
        {exelmans2016bedtime}
\bibfield{author}{\bibinfo{person}{Liese Exelmans} {and} \bibinfo{person}{Jan
  Van~den Bulck}.} \bibinfo{year}{2016}\natexlab{}.
\newblock \showarticletitle{Bedtime mobile phone use and sleep in adults}.
\newblock \bibinfo{journal}{\emph{Social Science \& Medicine}}
  \bibinfo{volume}{148} (\bibinfo{year}{2016}), \bibinfo{pages}{93--101}.
\newblock


\bibitem[\protect\citeauthoryear{Fernandez-Mendoza, He, Calhoun, Vgontzas,
  Liao, and Bixler}{Fernandez-Mendoza et~al\mbox{.}}{2020}]%
        {fernandez2020objective}
\bibfield{author}{\bibinfo{person}{Julio Fernandez-Mendoza},
  \bibinfo{person}{Fan He}, \bibinfo{person}{Susan~L Calhoun},
  \bibinfo{person}{Alexandros~N Vgontzas}, \bibinfo{person}{Duanping Liao},
  {and} \bibinfo{person}{Edward~O Bixler}.} \bibinfo{year}{2020}\natexlab{}.
\newblock \showarticletitle{Objective short sleep duration increases the risk
  of all-cause mortality associated with possible vascular cognitive
  impairment}.
\newblock \bibinfo{journal}{\emph{Sleep health}} \bibinfo{volume}{6},
  \bibinfo{number}{1} (\bibinfo{year}{2020}), \bibinfo{pages}{71--78}.
\newblock


\bibitem[\protect\citeauthoryear{Gu, Shangguan, Yang, and Liu}{Gu
  et~al\mbox{.}}{2015}]%
        {gu2015sleep}
\bibfield{author}{\bibinfo{person}{Weixi Gu}, \bibinfo{person}{Longfei
  Shangguan}, \bibinfo{person}{Zheng Yang}, {and} \bibinfo{person}{Yunhao
  Liu}.} \bibinfo{year}{2015}\natexlab{}.
\newblock \showarticletitle{Sleep hunter: Towards fine grained sleep stage
  tracking with smartphones}.
\newblock \bibinfo{journal}{\emph{IEEE Transactions on Mobile Computing}}
  \bibinfo{volume}{15}, \bibinfo{number}{6} (\bibinfo{year}{2015}),
  \bibinfo{pages}{1514--1527}.
\newblock


\bibitem[\protect\citeauthoryear{Hale, Troxel, and Buysse}{Hale
  et~al\mbox{.}}{2020}]%
        {hale2020sleep}
\bibfield{author}{\bibinfo{person}{Lauren Hale}, \bibinfo{person}{Wendy
  Troxel}, {and} \bibinfo{person}{Daniel~J Buysse}.}
  \bibinfo{year}{2020}\natexlab{}.
\newblock \showarticletitle{Sleep health: An opportunity for public health to
  address health equity}.
\newblock \bibinfo{journal}{\emph{Annual review of public health}}
  \bibinfo{volume}{41} (\bibinfo{year}{2020}), \bibinfo{pages}{81--99}.
\newblock


\bibitem[\protect\citeauthoryear{Hao, Xing, and Zhou}{Hao
  et~al\mbox{.}}{2013}]%
        {hao2013isleep}
\bibfield{author}{\bibinfo{person}{Tian Hao}, \bibinfo{person}{Guoliang Xing},
  {and} \bibinfo{person}{Gang Zhou}.} \bibinfo{year}{2013}\natexlab{}.
\newblock \showarticletitle{isleep: Unobtrusive sleep quality monitoring using
  smartphones}. In \bibinfo{booktitle}{\emph{Proceedings of the 11th ACM
  Conference on Embedded Networked Sensor Systems}}. \bibinfo{pages}{1--14}.
\newblock


\bibitem[\protect\citeauthoryear{He, Tu, Xiao, Su, and Tang}{He
  et~al\mbox{.}}{2020}]%
        {he2020effect}
\bibfield{author}{\bibinfo{person}{Jing-wen He}, \bibinfo{person}{Zhi-hao Tu},
  \bibinfo{person}{Lei Xiao}, \bibinfo{person}{Tong Su}, {and}
  \bibinfo{person}{Yun-xiang Tang}.} \bibinfo{year}{2020}\natexlab{}.
\newblock \showarticletitle{Effect of restricting bedtime mobile phone use on
  sleep, arousal, mood, and working memory: a randomized pilot trial}.
\newblock \bibinfo{journal}{\emph{PloS one}} \bibinfo{volume}{15},
  \bibinfo{number}{2} (\bibinfo{year}{2020}), \bibinfo{pages}{e0228756}.
\newblock


\bibitem[\protect\citeauthoryear{Hirshkowitz, Whiton, Albert, Alessi, Bruni,
  DonCarlos, Hazen, Herman, Katz, Kheirandish-Gozal, et~al\mbox{.}}{Hirshkowitz
  et~al\mbox{.}}{2015}]%
        {hirshkowitz2015national}
\bibfield{author}{\bibinfo{person}{Max Hirshkowitz}, \bibinfo{person}{Kaitlyn
  Whiton}, \bibinfo{person}{Steven~M Albert}, \bibinfo{person}{Cathy Alessi},
  \bibinfo{person}{Oliviero Bruni}, \bibinfo{person}{Lydia DonCarlos},
  \bibinfo{person}{Nancy Hazen}, \bibinfo{person}{John Herman},
  \bibinfo{person}{Eliot~S Katz}, \bibinfo{person}{Leila Kheirandish-Gozal},
  {et~al\mbox{.}}} \bibinfo{year}{2015}\natexlab{}.
\newblock \showarticletitle{National Sleep Foundation's sleep time duration
  recommendations: methodology and results summary}.
\newblock \bibinfo{journal}{\emph{Sleep health}} \bibinfo{volume}{1},
  \bibinfo{number}{1} (\bibinfo{year}{2015}), \bibinfo{pages}{40--43}.
\newblock


\bibitem[\protect\citeauthoryear{Hsu, Ahuja, Yue, Hristov, Kabelac, and
  Katabi}{Hsu et~al\mbox{.}}{2017}]%
        {hsu2017zero}
\bibfield{author}{\bibinfo{person}{Chen-Yu Hsu}, \bibinfo{person}{Aayush
  Ahuja}, \bibinfo{person}{Shichao Yue}, \bibinfo{person}{Rumen Hristov},
  \bibinfo{person}{Zachary Kabelac}, {and} \bibinfo{person}{Dina Katabi}.}
  \bibinfo{year}{2017}\natexlab{}.
\newblock \showarticletitle{Zero-effort in-home sleep and insomnia monitoring
  using radio signals}.
\newblock \bibinfo{journal}{\emph{Proceedings of the ACM on Interactive,
  mobile, wearable and ubiquitous technologies}} \bibinfo{volume}{1},
  \bibinfo{number}{3} (\bibinfo{year}{2017}), \bibinfo{pages}{1--18}.
\newblock


\bibitem[\protect\citeauthoryear{Huang, Wang, and de~Haan}{Huang
  et~al\mbox{.}}{2021}]%
        {huang2021nose}
\bibfield{author}{\bibinfo{person}{Zhengjie Huang}, \bibinfo{person}{Wenjin
  Wang}, {and} \bibinfo{person}{Gerard de Haan}.}
  \bibinfo{year}{2021}\natexlab{}.
\newblock \showarticletitle{Nose Breathing or Mouth Breathing? A
  Thermography-Based New Measurement for Sleep Monitoring}. In
  \bibinfo{booktitle}{\emph{Proceedings of the IEEE/CVF Conference on Computer
  Vision and Pattern Recognition}}. \bibinfo{pages}{3882--3888}.
\newblock


\bibitem[\protect\citeauthoryear{Kubiszewski, Fontaine, Rusch, and
  Hazouard}{Kubiszewski et~al\mbox{.}}{2014}]%
        {kubiszewski2014association}
\bibfield{author}{\bibinfo{person}{Violaine Kubiszewski},
  \bibinfo{person}{Roger Fontaine}, \bibinfo{person}{Emmanuel Rusch}, {and}
  \bibinfo{person}{Eric Hazouard}.} \bibinfo{year}{2014}\natexlab{}.
\newblock \showarticletitle{Association between electronic media use and sleep
  habits: An eight-day follow-up study}.
\newblock \bibinfo{journal}{\emph{International Journal of Adolescence and
  Youth}} \bibinfo{volume}{19}, \bibinfo{number}{3} (\bibinfo{year}{2014}),
  \bibinfo{pages}{395--407}.
\newblock


\bibitem[\protect\citeauthoryear{Kushida}{Kushida}{2012}]%
        {kushida2012encyclopedia}
\bibfield{author}{\bibinfo{person}{Clete Kushida}.}
  \bibinfo{year}{2012}\natexlab{}.
\newblock \bibinfo{booktitle}{\emph{Encyclopedia of sleep}}.
\newblock \bibinfo{publisher}{Academic Press}.
\newblock


\bibitem[\protect\citeauthoryear{Lauderdale, Knutson, Yan, Rathouz, Hulley,
  Sidney, and Liu}{Lauderdale et~al\mbox{.}}{2006}]%
        {lauderdale2006objectively}
\bibfield{author}{\bibinfo{person}{Diane~S Lauderdale},
  \bibinfo{person}{Kristen~L Knutson}, \bibinfo{person}{Lijing~L Yan},
  \bibinfo{person}{Paul~J Rathouz}, \bibinfo{person}{Stephen~B Hulley},
  \bibinfo{person}{Steve Sidney}, {and} \bibinfo{person}{Kiang Liu}.}
  \bibinfo{year}{2006}\natexlab{}.
\newblock \showarticletitle{Objectively measured sleep characteristics among
  early-middle-aged adults: the CARDIA study}.
\newblock \bibinfo{journal}{\emph{American journal of epidemiology}}
  \bibinfo{volume}{164}, \bibinfo{number}{1} (\bibinfo{year}{2006}),
  \bibinfo{pages}{5--16}.
\newblock


\bibitem[\protect\citeauthoryear{Magee and Hale}{Magee and Hale}{2012}]%
        {magee2012longitudinal}
\bibfield{author}{\bibinfo{person}{Lorrie Magee} {and} \bibinfo{person}{Lauren
  Hale}.} \bibinfo{year}{2012}\natexlab{}.
\newblock \showarticletitle{Longitudinal associations between sleep duration
  and subsequent weight gain: a systematic review}.
\newblock \bibinfo{journal}{\emph{Sleep medicine reviews}}
  \bibinfo{volume}{16}, \bibinfo{number}{3} (\bibinfo{year}{2012}),
  \bibinfo{pages}{231--241}.
\newblock


\bibitem[\protect\citeauthoryear{Mammen, Zakaria, Molom-Ochir, Trivedi, Shenoy,
  and Balan}{Mammen et~al\mbox{.}}{2021}]%
        {mammen2021wisleep}
\bibfield{author}{\bibinfo{person}{Priyanka~Mary Mammen},
  \bibinfo{person}{Camellia Zakaria}, \bibinfo{person}{Tergel Molom-Ochir},
  \bibinfo{person}{Amee Trivedi}, \bibinfo{person}{Prashant Shenoy}, {and}
  \bibinfo{person}{Rajesh Balan}.} \bibinfo{year}{2021}\natexlab{}.
\newblock \bibinfo{title}{WiSleep: Inferring Sleep Duration at Scale Using
  Passive WiFi Sensing}.
\newblock
\newblock
\showeprint[arxiv]{2102.03690}


\bibitem[\protect\citeauthoryear{Massar, Chua, Soon, Ng, Ong, Chee, Lee, Ghosh,
  and Chee}{Massar et~al\mbox{.}}{2021}]%
        {massar2021trait}
\bibfield{author}{\bibinfo{person}{Stijn~AA Massar}, \bibinfo{person}{Xin~Yu
  Chua}, \bibinfo{person}{Chun~Siong Soon}, \bibinfo{person}{Alyssa~SC Ng},
  \bibinfo{person}{Ju~Lynn Ong}, \bibinfo{person}{Nicholas~IYN Chee},
  \bibinfo{person}{Tih~Shih Lee}, \bibinfo{person}{Arko Ghosh}, {and}
  \bibinfo{person}{Michael~WL Chee}.} \bibinfo{year}{2021}\natexlab{}.
\newblock \showarticletitle{Trait-like nocturnal sleep behavior identified by
  combining wearable, phone-use, and self-report data}.
\newblock \bibinfo{journal}{\emph{NPJ digital medicine}} \bibinfo{volume}{4},
  \bibinfo{number}{1} (\bibinfo{year}{2021}), \bibinfo{pages}{1--10}.
\newblock


\bibitem[\protect\citeauthoryear{Mecacci and Zani}{Mecacci and Zani}{1983}]%
        {mecacci1983morningness}
\bibfield{author}{\bibinfo{person}{Luciano Mecacci} {and}
  \bibinfo{person}{Alberto Zani}.} \bibinfo{year}{1983}\natexlab{}.
\newblock \showarticletitle{Morningness-eveningness preferences and
  sleep-waking diary data of morning and evening types in student and worker
  samples}.
\newblock \bibinfo{journal}{\emph{Ergonomics}} \bibinfo{volume}{26},
  \bibinfo{number}{12} (\bibinfo{year}{1983}), \bibinfo{pages}{1147--1153}.
\newblock


\bibitem[\protect\citeauthoryear{Min, Doryab, Wiese, Amini, Zimmerman, and
  Hong}{Min et~al\mbox{.}}{2014}]%
        {min2014toss}
\bibfield{author}{\bibinfo{person}{Jun-Ki Min}, \bibinfo{person}{Afsaneh
  Doryab}, \bibinfo{person}{Jason Wiese}, \bibinfo{person}{Shahriyar Amini},
  \bibinfo{person}{John Zimmerman}, {and} \bibinfo{person}{Jason~I Hong}.}
  \bibinfo{year}{2014}\natexlab{}.
\newblock \showarticletitle{Toss`n'turn: smartphone as sleep and sleep quality
  detector}. In \bibinfo{booktitle}{\emph{Proceedings of the SIGCHI conference
  on human factors in computing systems}}. \bibinfo{pages}{477--486}.
\newblock


\bibitem[\protect\citeauthoryear{Mitler, Dement, and Dinges}{Mitler
  et~al\mbox{.}}{2000}]%
        {mitler2000sleep}
\bibfield{author}{\bibinfo{person}{Merrill~M Mitler},
  \bibinfo{person}{William~C Dement}, {and} \bibinfo{person}{David~F Dinges}.}
  \bibinfo{year}{2000}\natexlab{}.
\newblock \showarticletitle{Sleep medicine, public policy, and public health}.
\newblock \bibinfo{journal}{\emph{Principles and practice of sleep medicine}}
  \bibinfo{volume}{2}, \bibinfo{number}{4} (\bibinfo{year}{2000}),
  \bibinfo{pages}{453--462}.
\newblock


\bibitem[\protect\citeauthoryear{Networks}{Networks}{[n.d.]}]%
        {RTLS}
\bibfield{author}{\bibinfo{person}{A. Networks}.}
  \bibinfo{year}{[n.d.]}\natexlab{}.
\newblock \bibinfo{title}{RTLS - integrating with the RTLS data feed}.
\newblock
  \bibinfo{howpublished}{\url{https://community.arubanetworks.com/aruba/attachments/aruba/
  unified-wired-wireless-access/23715/1/RTLS integrationv6.docx}}.
\newblock
\newblock
\shownote{Accessed: 2021-11-07.}


\bibitem[\protect\citeauthoryear{Netzer, Stoohs, Netzer, Clark, and
  Strohl}{Netzer et~al\mbox{.}}{1999}]%
        {netzer1999using}
\bibfield{author}{\bibinfo{person}{Nikolaus~C Netzer},
  \bibinfo{person}{Riccardo~A Stoohs}, \bibinfo{person}{Cordula~M Netzer},
  \bibinfo{person}{Kathryn Clark}, {and} \bibinfo{person}{Kingman~P Strohl}.}
  \bibinfo{year}{1999}\natexlab{}.
\newblock \showarticletitle{Using the Berlin Questionnaire to identify patients
  at risk for the sleep apnea syndrome}.
\newblock \bibinfo{journal}{\emph{Annals of internal medicine}}
  \bibinfo{volume}{131}, \bibinfo{number}{7} (\bibinfo{year}{1999}),
  \bibinfo{pages}{485--491}.
\newblock


\bibitem[\protect\citeauthoryear{Ohayon, Reynolds~III, and Dauvilliers}{Ohayon
  et~al\mbox{.}}{2013}]%
        {ohayon2013excessive}
\bibfield{author}{\bibinfo{person}{Maurice~M Ohayon},
  \bibinfo{person}{Charles~F Reynolds~III}, {and} \bibinfo{person}{Yves
  Dauvilliers}.} \bibinfo{year}{2013}\natexlab{}.
\newblock \showarticletitle{Excessive sleep duration and quality of life}.
\newblock \bibinfo{journal}{\emph{Annals of neurology}} \bibinfo{volume}{73},
  \bibinfo{number}{6} (\bibinfo{year}{2013}), \bibinfo{pages}{785--794}.
\newblock


\bibitem[\protect\citeauthoryear{{Oura Health Oy}}{{Oura Health Oy}}{[n.d.]}]%
        {oura}
\bibfield{author}{\bibinfo{person}{{Oura Health Oy}}.}
  \bibinfo{year}{[n.d.]}\natexlab{}.
\newblock \bibinfo{title}{New Oura Ring Generation 3}.
\newblock
\newblock
\urldef\tempurl%
\url{https://ouraring.com}
\showURL{%
\tempurl}


\bibitem[\protect\citeauthoryear{Pedregosa, Varoquaux, Gramfort, Michel,
  Thirion, Grisel, Blondel, Prettenhofer, Weiss, Dubourg,
  et~al\mbox{.}}{Pedregosa et~al\mbox{.}}{2011}]%
        {pedregosa2011scikit}
\bibfield{author}{\bibinfo{person}{Fabian Pedregosa}, \bibinfo{person}{Ga{\"e}l
  Varoquaux}, \bibinfo{person}{Alexandre Gramfort}, \bibinfo{person}{Vincent
  Michel}, \bibinfo{person}{Bertrand Thirion}, \bibinfo{person}{Olivier
  Grisel}, \bibinfo{person}{Mathieu Blondel}, \bibinfo{person}{Peter
  Prettenhofer}, \bibinfo{person}{Ron Weiss}, \bibinfo{person}{Vincent
  Dubourg}, {et~al\mbox{.}}} \bibinfo{year}{2011}\natexlab{}.
\newblock \showarticletitle{Scikit-learn: Machine learning in Python}.
\newblock \bibinfo{journal}{\emph{Journal of machine learning research}}
  \bibinfo{volume}{12}, \bibinfo{number}{Oct} (\bibinfo{year}{2011}),
  \bibinfo{pages}{2825--2830}.
\newblock


\bibitem[\protect\citeauthoryear{Perry, Patil, and Presley-Cantrell}{Perry
  et~al\mbox{.}}{2013}]%
        {perry2013raising}
\bibfield{author}{\bibinfo{person}{Geraldine~S Perry},
  \bibinfo{person}{Susheel~P Patil}, {and} \bibinfo{person}{Letitia~R
  Presley-Cantrell}.} \bibinfo{year}{2013}\natexlab{}.
\newblock \showarticletitle{Raising awareness of sleep as a healthy behavior}.
\newblock \bibinfo{journal}{\emph{Preventing chronic disease}}
  \bibinfo{volume}{10} (\bibinfo{year}{2013}).
\newblock


\bibitem[\protect\citeauthoryear{{Pew Research Center}}{{Pew Research
  Center}}{2021b}]%
        {pewHomeBroadband}
\bibfield{author}{\bibinfo{person}{{Pew Research Center}}.}
  \bibinfo{year}{April 7, 2021}\natexlab{b}.
\newblock \bibinfo{title}{Internet/Broadband Fact Sheet}.
\newblock
\newblock
\urldef\tempurl%
\url{https://www.pewresearch.org/internet/fact-sheet/internet-broadband/#who-has-home-broadband}
\showURL{%
\tempurl}


\bibitem[\protect\citeauthoryear{{Pew Research Center}}{{Pew Research
  Center}}{2021c}]%
        {pewDeviceOwnership}
\bibfield{author}{\bibinfo{person}{{Pew Research Center}}.}
  \bibinfo{year}{April 7, 2021}\natexlab{c}.
\newblock \bibinfo{title}{Mobile Fact Sheet}.
\newblock
\newblock
\urldef\tempurl%
\url{https://www.pewresearch.org/internet/fact-sheet/mobile/#ownership-of-other-devices}
\showURL{%
\tempurl}


\bibitem[\protect\citeauthoryear{{Pew Research Center}}{{Pew Research
  Center}}{2020b}]%
        {pewPromisePitfall}
\bibfield{author}{\bibinfo{person}{{Pew Research Center}}.}
  \bibinfo{year}{December 8, 2020}\natexlab{b}.
\newblock \bibinfo{title}{The promise and pitfalls of using passive data to
  measure online news consumption}.
\newblock
\newblock
\urldef\tempurl%
\url{https://www.pewresearch.org/journalism/2020/12/08/the-promise-and-pitfalls-of-using-passive-data-to-measure-online-news-consumption/}
\showURL{%
\tempurl}


\bibitem[\protect\citeauthoryear{{Pew Research Center}}{{Pew Research
  Center}}{2021a}]%
        {pewOwnershipRace}
\bibfield{author}{\bibinfo{person}{{Pew Research Center}}.} \bibinfo{year}{July
  16, 2021}\natexlab{a}.
\newblock \bibinfo{title}{Home broadband adoption, computer ownership vary by
  race, ethnicity in the U.S.}
\newblock
\newblock
\urldef\tempurl%
\url{https://www.pewresearch.org/fact-tank/2021/07/16/home-broadband-adoption-computer-ownership-vary-by-race-ethnicity-in-the-u-s/}
\showURL{%
\tempurl}


\bibitem[\protect\citeauthoryear{{Pew Research Center}}{{Pew Research
  Center}}{2020a}]%
        {pewMoreDigital}
\bibfield{author}{\bibinfo{person}{{Pew Research Center}}.} \bibinfo{year}{June
  3, 2020}\natexlab{a}.
\newblock \bibinfo{title}{Experts Predict More Digital Innovation by 2030 Aimed
  at Enhancing Democracy}.
\newblock
\newblock
\urldef\tempurl%
\url{https://www.pewresearch.org/internet/2020/06/30/experts-predict-more-digital-innovation-by-2030-aimed-at-enhancing-democracy/}
\showURL{%
\tempurl}


\bibitem[\protect\citeauthoryear{Phillips, Clerx, O'Brien, Sano, Barger,
  Picard, Lockley, Klerman, and Czeisler}{Phillips et~al\mbox{.}}{2017}]%
        {phillips2017irregular}
\bibfield{author}{\bibinfo{person}{Andrew~JK Phillips},
  \bibinfo{person}{William~M Clerx}, \bibinfo{person}{Conor~S O'Brien},
  \bibinfo{person}{Akane Sano}, \bibinfo{person}{Laura~K Barger},
  \bibinfo{person}{Rosalind~W Picard}, \bibinfo{person}{Steven~W Lockley},
  \bibinfo{person}{Elizabeth~B Klerman}, {and} \bibinfo{person}{Charles~A
  Czeisler}.} \bibinfo{year}{2017}\natexlab{}.
\newblock \showarticletitle{Irregular sleep/wake patterns are associated with
  poorer academic performance and delayed circadian and sleep/wake timing}.
\newblock \bibinfo{journal}{\emph{Scientific reports}} \bibinfo{volume}{7},
  \bibinfo{number}{1} (\bibinfo{year}{2017}), \bibinfo{pages}{1--13}.
\newblock


\bibitem[\protect\citeauthoryear{Polimis, Rokem, and Hazelton}{Polimis
  et~al\mbox{.}}{2017}]%
        {polimisconfidence}
\bibfield{author}{\bibinfo{person}{Kivan Polimis}, \bibinfo{person}{Ariel
  Rokem}, {and} \bibinfo{person}{Bryna Hazelton}.}
  \bibinfo{year}{2017}\natexlab{}.
\newblock \showarticletitle{Confidence Intervals for Random Forests in Python}.
\newblock \bibinfo{journal}{\emph{Journal of Open Source Software}}
  \bibinfo{volume}{2}, \bibinfo{number}{1} (\bibinfo{year}{2017}).
\newblock


\bibitem[\protect\citeauthoryear{Press and Teukolsky}{Press and
  Teukolsky}{1990}]%
        {press1990savitzky}
\bibfield{author}{\bibinfo{person}{William~H Press} {and}
  \bibinfo{person}{Saul~A Teukolsky}.} \bibinfo{year}{1990}\natexlab{}.
\newblock \showarticletitle{Savitzky-Golay smoothing filters}.
\newblock \bibinfo{journal}{\emph{Computers in Physics}} \bibinfo{volume}{4},
  \bibinfo{number}{6} (\bibinfo{year}{1990}), \bibinfo{pages}{669--672}.
\newblock


\bibitem[\protect\citeauthoryear{Rahman, Adams, Ravichandran, Zhang, Patel,
  Kientz, and Choudhury}{Rahman et~al\mbox{.}}{2015}]%
        {rahman2015dopplesleep}
\bibfield{author}{\bibinfo{person}{Tauhidur Rahman},
  \bibinfo{person}{Alexander~T Adams}, \bibinfo{person}{Ruth~Vinisha
  Ravichandran}, \bibinfo{person}{Mi Zhang}, \bibinfo{person}{Shwetak~N Patel},
  \bibinfo{person}{Julie~A Kientz}, {and} \bibinfo{person}{Tanzeem Choudhury}.}
  \bibinfo{year}{2015}\natexlab{}.
\newblock \showarticletitle{Dopplesleep: A contactless unobtrusive sleep
  sensing system using short-range doppler radar}. In
  \bibinfo{booktitle}{\emph{Proceedings of the 2015 ACM International Joint
  Conference on Pervasive and Ubiquitous Computing}}. \bibinfo{pages}{39--50}.
\newblock


\bibitem[\protect\citeauthoryear{Rosenberger, Buman, Haskell, McConnell, and
  Carstensen}{Rosenberger et~al\mbox{.}}{2016}]%
        {rosenberger201624}
\bibfield{author}{\bibinfo{person}{Mary~E Rosenberger},
  \bibinfo{person}{Matthew~P Buman}, \bibinfo{person}{William~L Haskell},
  \bibinfo{person}{Michael~V McConnell}, {and} \bibinfo{person}{Laura~L
  Carstensen}.} \bibinfo{year}{2016}\natexlab{}.
\newblock \showarticletitle{24 hours of sleep, sedentary behavior, and physical
  activity with nine wearable devices}.
\newblock \bibinfo{journal}{\emph{Medicine and science in sports and exercise}}
  \bibinfo{volume}{48}, \bibinfo{number}{3} (\bibinfo{year}{2016}),
  \bibinfo{pages}{457}.
\newblock


\bibitem[\protect\citeauthoryear{Smaldone, Honig, and Byrne}{Smaldone
  et~al\mbox{.}}{2007}]%
        {smaldone2007sleepless}
\bibfield{author}{\bibinfo{person}{Arlene Smaldone}, \bibinfo{person}{Judy~C
  Honig}, {and} \bibinfo{person}{Mary~W Byrne}.}
  \bibinfo{year}{2007}\natexlab{}.
\newblock \showarticletitle{Sleepless in America: inadequate sleep and
  relationships to health and well-being of our nation's children}.
\newblock \bibinfo{journal}{\emph{Pediatrics}} \bibinfo{volume}{119},
  \bibinfo{number}{Supplement\_1} (\bibinfo{year}{2007}),
  \bibinfo{pages}{S29--S37}.
\newblock


\bibitem[\protect\citeauthoryear{Steer and Beck}{Steer and Beck}{1997}]%
        {steer1997beck}
\bibfield{author}{\bibinfo{person}{Robert~A Steer} {and}
  \bibinfo{person}{Aaron~T Beck}.} \bibinfo{year}{1997}\natexlab{}.
\newblock \showarticletitle{Beck Anxiety Inventory.}
\newblock  (\bibinfo{year}{1997}).
\newblock


\bibitem[\protect\citeauthoryear{Steer, Brown, Beck, and Sanderson}{Steer
  et~al\mbox{.}}{2001}]%
        {steer2001mean}
\bibfield{author}{\bibinfo{person}{Robert~A Steer}, \bibinfo{person}{Gregory~K
  Brown}, \bibinfo{person}{Aaron~T Beck}, {and} \bibinfo{person}{William~C
  Sanderson}.} \bibinfo{year}{2001}\natexlab{}.
\newblock \showarticletitle{Mean Beck Depression Inventory--II scores by
  severity of major depressive episode}.
\newblock \bibinfo{journal}{\emph{Psychological reports}} \bibinfo{volume}{88},
  \bibinfo{number}{3\_suppl} (\bibinfo{year}{2001}),
  \bibinfo{pages}{1075--1076}.
\newblock


\bibitem[\protect\citeauthoryear{Taub}{Taub}{1978}]%
        {taub1978behavioral}
\bibfield{author}{\bibinfo{person}{John~M Taub}.}
  \bibinfo{year}{1978}\natexlab{}.
\newblock \showarticletitle{Behavioral and psychophysiological correlates of
  irregularity in chronic sleep routines}.
\newblock \bibinfo{journal}{\emph{Biological Psychology}} \bibinfo{volume}{7},
  \bibinfo{number}{1-2} (\bibinfo{year}{1978}), \bibinfo{pages}{37--53}.
\newblock


\bibitem[\protect\citeauthoryear{Tiron, Lyon, Kilroy, Osman, Kelly, O'Mahony,
  Lopes, Coffey, McMahon, Wren, et~al\mbox{.}}{Tiron et~al\mbox{.}}{2020}]%
        {tiron2020screening}
\bibfield{author}{\bibinfo{person}{Roxana Tiron}, \bibinfo{person}{Graeme
  Lyon}, \bibinfo{person}{Hannah Kilroy}, \bibinfo{person}{Ahmed Osman},
  \bibinfo{person}{Nicola Kelly}, \bibinfo{person}{Niall O'Mahony},
  \bibinfo{person}{Cesar Lopes}, \bibinfo{person}{Sam Coffey},
  \bibinfo{person}{Stephen McMahon}, \bibinfo{person}{Michael Wren},
  {et~al\mbox{.}}} \bibinfo{year}{2020}\natexlab{}.
\newblock \showarticletitle{Screening for obstructive sleep apnea with novel
  hybrid acoustic smartphone app technology}.
\newblock \bibinfo{journal}{\emph{Journal of Thoracic Disease}}
  \bibinfo{volume}{12}, \bibinfo{number}{8} (\bibinfo{year}{2020}),
  \bibinfo{pages}{4476}.
\newblock


\bibitem[\protect\citeauthoryear{Toh, Howie, Coenen, and Straker}{Toh
  et~al\mbox{.}}{2019}]%
        {toh2019moment}
\bibfield{author}{\bibinfo{person}{Siao~Hui Toh}, \bibinfo{person}{Erin~K
  Howie}, \bibinfo{person}{Pieter Coenen}, {and} \bibinfo{person}{Leon~M
  Straker}.} \bibinfo{year}{2019}\natexlab{}.
\newblock \showarticletitle{``From the moment I wake up I will use it... every
  day, very hour'': a qualitative study on the patterns of adolescents' mobile
  touch screen device use from adolescent and parent perspectives}.
\newblock \bibinfo{journal}{\emph{BMC pediatrics}} \bibinfo{volume}{19},
  \bibinfo{number}{1} (\bibinfo{year}{2019}), \bibinfo{pages}{1--16}.
\newblock


\bibitem[\protect\citeauthoryear{Trivedi, Gummeson, and Shenoy}{Trivedi
  et~al\mbox{.}}{2020}]%
        {trivedi2020empirical}
\bibfield{author}{\bibinfo{person}{Amee Trivedi}, \bibinfo{person}{Jeremy
  Gummeson}, {and} \bibinfo{person}{Prashant Shenoy}.}
  \bibinfo{year}{2020}\natexlab{}.
\newblock \showarticletitle{Empirical characterization of mobility of
  multi-device internet users}.
\newblock \bibinfo{journal}{\emph{arXiv preprint arXiv:2003.08512}}
  (\bibinfo{year}{2020}).
\newblock


\bibitem[\protect\citeauthoryear{Umematsu, Sano, Taylor, and Picard}{Umematsu
  et~al\mbox{.}}{2019}]%
        {umematsu2019improving}
\bibfield{author}{\bibinfo{person}{Terumi Umematsu}, \bibinfo{person}{Akane
  Sano}, \bibinfo{person}{Sara Taylor}, {and} \bibinfo{person}{Rosalind~W
  Picard}.} \bibinfo{year}{2019}\natexlab{}.
\newblock \showarticletitle{Improving students' daily life stress forecasting
  using LSTM neural networks}. In \bibinfo{booktitle}{\emph{2019 IEEE EMBS
  International Conference on Biomedical \& Health Informatics (BHI)}}. IEEE,
  \bibinfo{pages}{1--4}.
\newblock


\bibitem[\protect\citeauthoryear{Van~Deursen, Bolle, Hegner, and
  Kommers}{Van~Deursen et~al\mbox{.}}{2015}]%
        {van2015modeling}
\bibfield{author}{\bibinfo{person}{Alexander~JAM Van~Deursen},
  \bibinfo{person}{Colin~L Bolle}, \bibinfo{person}{Sabrina~M Hegner}, {and}
  \bibinfo{person}{Piet~AM Kommers}.} \bibinfo{year}{2015}\natexlab{}.
\newblock \showarticletitle{Modeling habitual and addictive smartphone
  behavior: The role of smartphone usage types, emotional intelligence, social
  stress, self-regulation, age, and gender}.
\newblock \bibinfo{journal}{\emph{Computers in human behavior}}
  \bibinfo{volume}{45} (\bibinfo{year}{2015}), \bibinfo{pages}{411--420}.
\newblock


\bibitem[\protect\citeauthoryear{Wager, Hastie, and Efron}{Wager
  et~al\mbox{.}}{2014}]%
        {wager2014confidence}
\bibfield{author}{\bibinfo{person}{Stefan Wager}, \bibinfo{person}{Trevor
  Hastie}, {and} \bibinfo{person}{Bradley Efron}.}
  \bibinfo{year}{2014}\natexlab{}.
\newblock \showarticletitle{Confidence intervals for random forests: The
  jackknife and the infinitesimal jackknife}.
\newblock \bibinfo{journal}{\emph{The Journal of Machine Learning Research}}
  \bibinfo{volume}{15}, \bibinfo{number}{1} (\bibinfo{year}{2014}),
  \bibinfo{pages}{1625--1651}.
\newblock


\bibitem[\protect\citeauthoryear{Yu, Wang, Niu, Zeng, Gu, Wang, Guan, and
  Zhang}{Yu et~al\mbox{.}}{2021}]%
        {yu2021wifi}
\bibfield{author}{\bibinfo{person}{Bohan Yu}, \bibinfo{person}{Yuxiang Wang},
  \bibinfo{person}{Kai Niu}, \bibinfo{person}{Youwei Zeng},
  \bibinfo{person}{Tao Gu}, \bibinfo{person}{Leye Wang},
  \bibinfo{person}{Cuntai Guan}, {and} \bibinfo{person}{Daqing Zhang}.}
  \bibinfo{year}{2021}\natexlab{}.
\newblock \showarticletitle{WiFi-Sleep: Sleep Stage Monitoring Using Commodity
  Wi-Fi Devices}.
\newblock \bibinfo{journal}{\emph{IEEE Internet of Things Journal}}
  (\bibinfo{year}{2021}).
\newblock


\bibitem[\protect\citeauthoryear{Yue, Yang, Wang, Rahul, and Katabi}{Yue
  et~al\mbox{.}}{2020}]%
        {yue2020bodycompass}
\bibfield{author}{\bibinfo{person}{Shichao Yue}, \bibinfo{person}{Yuzhe Yang},
  \bibinfo{person}{Hao Wang}, \bibinfo{person}{Hariharan Rahul}, {and}
  \bibinfo{person}{Dina Katabi}.} \bibinfo{year}{2020}\natexlab{}.
\newblock \showarticletitle{BodyCompass: Monitoring sleep posture with wireless
  signals}.
\newblock \bibinfo{journal}{\emph{Proceedings of the ACM on Interactive,
  Mobile, Wearable and Ubiquitous Technologies}} \bibinfo{volume}{4},
  \bibinfo{number}{2} (\bibinfo{year}{2020}), \bibinfo{pages}{1--25}.
\newblock


\bibitem[\protect\citeauthoryear{Zhai, Perez-Pozuelo, Clifton, Palotti, and
  Guan}{Zhai et~al\mbox{.}}{2020}]%
        {zhai2020making}
\bibfield{author}{\bibinfo{person}{Bing Zhai}, \bibinfo{person}{Ignacio
  Perez-Pozuelo}, \bibinfo{person}{Emma~AD Clifton}, \bibinfo{person}{Joao
  Palotti}, {and} \bibinfo{person}{Yu Guan}.} \bibinfo{year}{2020}\natexlab{}.
\newblock \showarticletitle{Making sense of sleep: Multimodal sleep stage
  classification in a large, diverse population using movement and cardiac
  sensing}.
\newblock \bibinfo{journal}{\emph{Proceedings of the ACM on Interactive,
  Mobile, Wearable and Ubiquitous Technologies}} \bibinfo{volume}{4},
  \bibinfo{number}{2} (\bibinfo{year}{2020}), \bibinfo{pages}{1--33}.
\newblock


\end{thebibliography}



\end{document}